\documentclass[10pt,
amsmath,
amssymb,
aps,
prd,
nofootinbib,
noeprint,
preprintnumbers,
superscriptaddress,
twocolumn]{revtex4-1}

\usepackage{graphicx}
\usepackage{amsmath}
\usepackage[toc]{appendix}
\usepackage{lineno}
\usepackage[dvipsnames]{xcolor}
\usepackage{soul}
\usepackage[utf8]{inputenc}
\usepackage[T1]{fontenc}
\usepackage{newtxtext,newtxmath}

\usepackage{graphicx}
\usepackage[%
    allcolors=blue,
    colorlinks=true,
]{hyperref}
\usepackage{orcidlink}

\newcommand{\GRASP}{\affiliation{Institute for Gravitational and Subatomic Physics (GRASP), Utrecht University, Princetonplein 1, 3584 CC Utrecht, The Netherlands}}
\newcommand{\nikhef}{\affiliation{Nikhef -- National Institute for Subatomic Physics, Science Park 105, 1098 XG Amsterdam, The Netherlands}}
\newcommand{\OU}{\affiliation{Faculty of Science, Open Universiteit, Valkenburgerweg 177, 6419 AT Heerlen, The Netherlands}}
\newcommand{\UCL}{\affiliation{Centre for Cosmology, Particle Physics and Phenomenology - CP3, Universit\'e Catholique de Louvain, Louvain-La-Neuve, B-1348, Belgium}}
\newcommand{\ORB}{\affiliation{Royal Observatory of Belgium, Avenue Circulaire, 3, 1180 Uccle, Belgium}}
\newcommand{\KUL}{\affiliation{Leuven Gravity Institute, KU Leuven, Celestijnenlaan 200D box 2415, 3001 Leuven, Belgium}}
\newcommand{\PHYKUL}{\affiliation{Department of Physics and Astronomy, Laboratory for Semiconductor Physics, KU Leuven, B-3001 Leuven, Belgium}}
\newcommand{\EEKUL}{\affiliation{KU Leuven, Department of Electrical Engineering (ESAT), STADIUS Center for Dynamical Systems, Signal Processing and Data Analytics, B-3001 Leuven, Belgium}}
\newcommand{\Upisa}{\affiliation{Dipartimento di Fisica “E. Fermi”, Universit\`{a} di Pisa, I-56127 Pisa, Italy}}

\begin{document}
\title{Null Stream Based Third-generation-ready Glitch Mitigation for 
Gravitational Wave Measurements} 
\author{Harsh~Narola}
\email{h.b.narola@uu.nl}
\GRASP \nikhef

\author{Thibeau~Wouters}
\GRASP \nikhef

\author{Luca~Negri}
\GRASP \nikhef

\author{Melissa~Lopez}
\GRASP \nikhef

\author{Tom~Dooney}
\OU

\author{Francesco~Cireddu}
\KUL \PHYKUL \Upisa

\author{Milan~Wils}
\KUL \PHYKUL

\author{Isaac~C.~F.~Wong}
\KUL \EEKUL

\author{Peter~T.~H.~Pang}
\nikhef \GRASP

\author{Justin~Janquart}
\UCL \ORB \GRASP \nikhef

\author{Anuradha~Samajdar}
\GRASP \nikhef

\author{Chris~Van~Den~Broeck}
\GRASP \nikhef

\author{Tjonnie~G.~F.~Li}
\KUL \PHYKUL \EEKUL

\date{\today}
\newcommand{\ETT}{ET-$\Delta$}
\begin{abstract}
Gravitational wave (GW) detectors routinely encounter transient noise bursts, known as glitches, 
which are caused by either instrumental or environmental factors. Due to their high occurrence rate, 
glitches can overlap with GW signals, as in the notable case of GW170817, the first detection of a 
binary neutron star merger. Accurate reconstruction and subtraction of these glitches is a 
challenging problem that must be addressed to ensure that scientific conclusions drawn from the data 
are reliable. This problem will exacerbate with third-generation detectors like Einstein Telescope 
(ET) due to their higher detection rates of GWs and the longer duration of signals within the 
sensitivity band of the detectors. Robust glitch mitigation algorithms are, therefore, crucial 
for maximizing the scientific output of next-generation GW detectors. 
For the first time, we demonstrate how the null stream inherent in ET's unique triangular configuration 
can be leveraged by state-of-the-art glitch mitigation methodology to essentially undo the effect of 
glitches for the purpose of estimating the parameters of the source. 
The null stream based approach enables mitigation and subtraction of glitches that occur 
arbitrarily close to the peak of the signal without any significant effect on the quality of parameter 
measurements, and achieves an order of magnitude computational speed-up compared to when the null 
stream is not available. By contrast, without the null stream, 
significant biases can occur in the glitch reconstruction, which deteriorate the quality of subsequent 
measurements of the source parameters. This demonstrates a clear edge which the null stream can offer 
for precision GW science in the ET era.
\end{abstract}

\maketitle

%\section{Introduction}
\section{Introduction} The detection of gravitational waves (GWs) by Advanced 
LIGO~\cite{LIGOScientific:2014pky} and Advanced Virgo~\cite{VIRGO:2014yos} has opened a new 
window for observing the Universe, leading to numerous 
discoveries~\cite{LIGOScientific:2021aug, KAGRA:2021duu, LIGOScientific:2021sio, KAGRA:2021duu, LIGOScientific:2023bwz}. 
Extracting scientific insights from GW data fundamentally relies on accurately estimating the 
source parameters~\cite{Ashton:2018jfp, Romero-Shaw:2020owr, Narola:2023men}. A typical parameter 
estimation procedure assumes the detector noise to be stationary and 
Gaussian~\cite{LIGOScientific:2019hgc}. However, a common scenario for this assumption to fail 
is the occurrence of transient noise bursts, known as \textit{glitches}~\cite{LIGOScientific:2017tza, LIGO:2021ppb, Virgo:2022fxr, LIGO:2024kkz}. They can originate from environmental sources (e.g., earthquakes, wind, anthropogenic noise) or from instrumental factors (e.g., control systems, electronic components~\cite{LIGO:2020zwl}), though their origins remain unknown in many cases~\cite{Cabero:2019orq}.

Glitches can bias parameter estimation and also corrupt subsequent analyses when they overlap with 
signals~\cite{Pankow:2018qpo, Payne:2022spz, Udall:2024ovp, Gupta:2024gun}. Of the 90 confident 
GW detections in GWTC-3~\cite{KAGRA:2021vkt}, $\simeq 20$ required some degree of glitch mitigation. Such scenarios 
are expected to become more common with the increase in the detection rate of GWs since glitches occur 
at a much higher rate ($\simeq 1$ per minute) compared to GW 
detections~\cite{LIGOScientific:2018mvr, LIGOScientific:2020ibl, KAGRA:2021vkt}. The binary neutron star event GW170817~\cite{LIGOScientific:2017vwq} is a notable example of a glitch-contaminated signal, 
featuring a loud glitch near its merger in LIGO-Livingston data, which was extensively analyzed for 
its impact on parameter estimates~\cite{Pankow:2018qpo, Chatziioannou:2021ezd}. 

Third-generation (3G) ground-based GW detectors, such as Einstein Telescope (ET)~\cite{Punturo:2010zz} 
and Cosmic Explorer (CE)~\cite{Evans:2021gyd}, are expected to be ten times more sensitive than current 
detectors and will have a wider frequency sensitivity 
band~\cite{Hild:2010id, Maggiore:2019uih, Reitze:2019iox}, resulting in higher detection rates and 
longer signal durations. If 3G detectors experience glitches at rates similar to current detectors 
(approximately one per minute during the third LIGO-Virgo observing 
run~\cite{LIGOScientific:2020ibl, KAGRA:2021vkt}) while continuously observing 
GW signals~\cite{Maggiore:2019uih, Branchesi:2023mws}, many signals will be corrupted by glitches. 
With a GW detection rate of one every five minutes~\cite{Maggiore:2019uih}, a glitch is expected within 
100 ms of a GW merger approximately once per day.

Most glitches exhibit no clear correlation with instrumental or environmental 
sources~\cite{VIRGO:2012oxz}, rendering their origins difficult to 
diagnose~\cite{Nuttall:2018xhi, Davis:2022dnd}. A challenging problem in 
reconstructing a glitch that overlaps with a GW signal is to avoid misidentifying a part of the 
glitch as signal, and vice versa~\cite{Chatziioannou:2021ezd, Hourihane:2022doe, Ghonge:2023ksb}. This 
problem is especially difficult when the signal model is known with limited 
accuracy~\cite{Buonanno:1998gg, Husa:2015iqa, Boyle:2019kee, Lopez:2024xby} and the glitch model remains 
largely unknown. The issue is expected to aggravate for 3G detectors, as GW modeling will become more 
complex due to the presence of long, loud, and overlapping 
signals~\cite{Samajdar:2021egv, Pizzati:2021apa, Antonelli:2021vwg, Janquart:2022fzz, Relton:2021cax}. 
Further complexiteis are added when the signal is challenging to model, e.g. GWs from core-collapse 
supernovae~\cite{LIGOScientific:2019ryq, LopezPortilla:2020odz, Szczepanczyk:2023ihe}, the 
post-merger phase of a binary neutron star signal~\cite{Dietrich:2020eud}, or GWs carrying imprints of 
microlensing, environmental effects, orbital eccentricity, or deviation from general 
relativity~\cite{Roy:2024rhe, Leong:2023nuk, Wright:2021cbn, Gupte:2024jfe}. The problem may worsen if we 
encounter new types of glitches as we collect more data.

Here, we show that the null stream inherent in the trinagular configuration of Einstein Telescope 
(ET-$\Delta$)~\cite{ESFRI} enables a 3G-ready method for glitch removal. The use of null stream 
fully prevents contamination of the signal due to overlapping glitches, leading to accurate source parameter 
measurements. In contrast, we show that the alternative design of ET -- a network of two L-shaped 
misaligned 
interferometers (ET-2L) which does not inherit the null stream~\cite{Branchesi:2023mws} --leads to 
incorrect measurements due to the contamination caused by glitches. This comparison highlights the 
importance of null stream and presents a way toward achieving the precision science goals of the 
3G era.\\
%\footnote{For the purposes of this paper we will take ET's design to be 
%an equilateral triangle composed of three V-shaped detectors with 10 km arm length , but 
%other configurations have also been investigated .}

\section{Glitches} Glitches exhibit a wide range of time-frequency morphologies and could hamper GW data 
analyses in various ways. They reduce analyzable data, elevate the noise floor, generate false positives, 
distort the detectors' noise power spectral density estimates, and decrease the significance of GW 
signal candidates~\cite{LIGOScientific:2017tza, Steltner:2023cfk, Steltner:2021qjy}. Additionally, they 
could bias astrophysical parameter estimates by complicating the separation between glitches and genuine 
GW signals~\cite{Pankow:2018qpo, Davis:2018yrz, LIGOScientific:2018kdd}. Consequently, understanding and 
mitigating noise sources from both instrumental and data analysis perspectives remains a major focus of 
the LIGO-Virgo-KAGRA Collaboration~\cite{LIGOScientific:2019hgc, LIGO:2021ppb, Virgo:2022fxr}.

Among the most problematic classes are ``blip'' glitches, which are short-lived bursts of power 
(with durations $\lesssim 0.2$ s) exhibiting a characteristic symmetric ``teardrop'' shape in the 
time-frequency domain, typically within the frequency range of [30, 250] Hz~\cite{Cabero:2019orq}. 
Thanks to recent developments in machine learning based methods, it is now possible to simulate glitches.  
For example, the \texttt{gengli} codebase can generate blip glitches akin to the ones observed during the second observing run 
of Advanced LIGO and Virgo~\cite{Lopez:2022lkd, Lopez:2022dho}.\\

\section{Null stream} The \textit{null stream} is a linear combination of data from a detector network that 
cancels GW signals~\cite{Guersel:1989th, Freise:2008dk, Sutton:2009gi, Pang:2020pfz, Wong:2021cmp, Goncharov:2022dgl}. 
Typically, constructing a null stream for a network of three detectors requires knowledge of the 
sky location of the GW source. However, the triangular configuration of ET-$\Delta$ uniquely inherits a 
sky position independent null stream, 
allowing it to be generated by summing the data from the 
three detectors.

We denote the individual detectors in ET-$\Delta$ as ${\rm ET}_1$, ${\rm ET}_2$, and ${\rm ET}_3$, with 
corresponding data streams $\vec{d}_1$, $\vec{d}_2$, and $\vec{d}_3$. Suppose that ${\rm ET}_1$ 
records a glitch $\vec{g}$ overlapping with a GW signal $\vec{h}$, while ${\rm ET}_2$ and ${\rm ET}_3$ 
contain only the GW signal and stationary, Gaussian noise. Then
\begin{equation}
\begin{aligned}
  \vec{d}_1 &= \vec{h}_1 + \vec{n}_1 + \vec{g},\\
  \vec{d}_2 &= \vec{h}_2 + \vec{n}_2,\\
  \vec{d}_3 &= \vec{h}_3 + \vec{n}_3,
\end{aligned}
\end{equation}
where $\vec{h}_i$ and $\vec{n}_i$ represent the signal projection and Gaussian noise in the 
$i^\mathrm{th}$ detector, respectively. As previously mentioned, due to the geometry of the detectors, 
$\sum_{i=1}^3 \vec{h}_i = 0$. The null stream $\vec{d}_{\rm null}$ is then given by 
\begin{align}
\label{eq:sum_data}
  \vec{d}_{\mathrm{null}} &\equiv \frac{1}{\sqrt{3}}\left(\vec{d}_1 + \vec{d}_2 + \vec{d}_3\right),\\
  \label{eq:no_signal_eq}
  &= \frac{1}{\sqrt{3}}\left(\vec{g} + \vec{n}_1 + \vec{n}_2 + \vec{n}_3\right) \nonumber \\
  &= \vec{g}_{\rm null} + \vec{n}_{\rm null},
\end{align}
where $\vec{n}_{\rm null} \equiv (\vec{n}_1 + \vec{n}_2 + \vec{n}_3)/\sqrt{3}$. The normalization factor $1/\sqrt{3}$ ensures that the noise's power spectral density equals the average of those of the individual detectors. Since the null stream contains no GW signal, it enables glitch reconstruction without the possibility of signal contamination.

\begin{figure}
    \centering
    \includegraphics[width=0.5\textwidth]{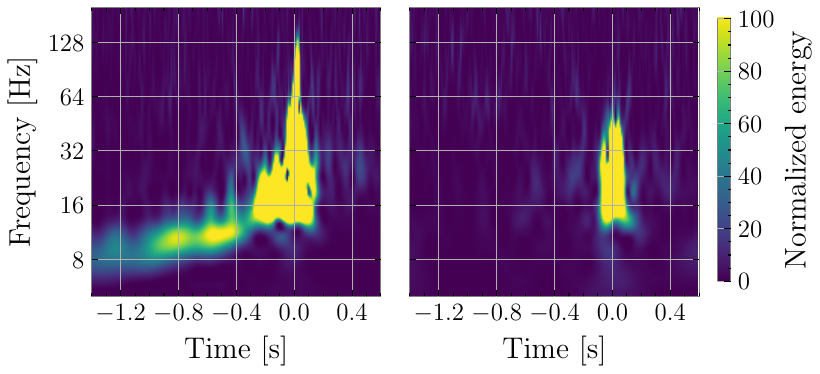}
    \caption{\textbf{Left:} A simulated ``blip'' glitch overlapping with a GW150914-like signal in the ${\rm ET}_1$ detector. The chirping contour represents the GW signal, while the vertical contour near the merger time indicates the glitch. \textbf{Right:} The null stream corresponding to the data segment shown on the left. Only the glitch is present since the signal does not contribute to the null stream.}
    \label{fig:glitch_panel}
\end{figure}

As shown in Fig.~\ref{fig:glitch_panel}, we can construct the null stream for a data segment using 
Eq.~\eqref{eq:sum_data}, resulting in a null stream spectrogram where the GW signal is absent, 
consistent with Eq.~\eqref{eq:no_signal_eq}.\\

\section{BayesWave} A commonly used method for glitch reconstruction is 
\texttt{BayesWave} \cite{Cornish:2014kda, Cornish:2020dwh, Chatziioannou:2021ezd, Hourihane:2022doe} 
which can perform simultaneous modeling of signal and glitch. Up to an irrelevant constant, 
the log-likelihood in the presence of Gaussian noise is given by \cite{Veitch:2009hd}
\begin{equation}
\begin{aligned}
    \log p(\vec{d} | \Theta, \Phi) = 
    -\frac{1}{2}\langle \vec{d}-\vec{g}(\Theta) - \vec{h}(\Phi) | \vec{d} - \vec{g}(\Theta) - \vec{h}(\Phi)\rangle,
\end{aligned}
\end{equation}
where $\vec{g}(\Theta)$ ($\vec{h}(\Phi)$) denotes the template of the glitch (signal) given its model 
parameters $\Theta$ ($\Phi$). The angular brackets denote the noise-weighted inner product
\begin{equation}
    \label{eq:innerp}
    \langle \vec{a}|\vec{b} \rangle = \frac{4}{T} \Re \sum_{f} \frac{\tilde{a}(f)\tilde{b}^*(f)}{S_n(f)},
\end{equation}
where $f$, $S_n(f)$, and $T$ are the frequency, one-sided noise power spectral density, and the data 
segment duration, respectively. $\Re$ specifies the real component, $\tilde{b}(f)$ denotes the Fourier 
transform of $b(t)$, and $\tilde{b}^*(f)$ its complex conjugate.

The glitch template $\vec{g}$ is a sum of sine-Gaussian wavelets, 
\begin{equation}
    g(t; \Theta) = \sum_{j=1}^{N} A_j  e^{-\frac{\left(t-t_{0, j}\right)^2}{\tau_{j}^2}} \sin{\left(2\pi f_{0, j} \left(t-t_{0, j}\right) + \phi_{0, j}\right)}
\end{equation}
where $\{A_j, t_{0, j}, f_{0, j}, \phi_{0, j}, \tau_j\}$ are the model parameters of the 
$j^{\mathrm{th}}$ wavelet, and the total number of wavelets $N$ is itself a model parameter.

As introduced in Ref.~\cite{Chatziioannou:2021ezd}, instead of modeling the GW signal itself with 
sine-Gaussian wavelets, a realistic signal waveform is employed with the expectation of reducing the mismodelling of the signal due to a glitch. 
The simultaneous modeling of signal and the glitch is done using a 
Reversible-Jump Markov Chain Monte Carlo method~\cite{Green:1995mxx} suitable for a 
variable-dimensional model such as this.\\

\section{Leveraging the null stream for glitch mitigation} For the ET-$\Delta$ configuration, 
instead of performing simultaneous modeling of signal and glitch, we can make use of the null stream. As shown in 
Eq.~\eqref{eq:no_signal_eq}, the signal is completely absent from $\vec{d}_{\rm null}$; hence we 
can use $\vec{d}_{\rm null}$ to reconstruct the glitch by itself, avoiding any possibility of signal 
contamination. The individual modeling of the glitch also reduces the model's dimensionality compared 
to simultaneous modeling. Up to a constant, the log-likelihood is now given by
\begin{equation}
\begin{aligned}
    \log p(\vec{d}_{\rm null} | \Theta) = -\frac{1}{2}\langle \vec{d}_{\rm null}-\vec{g}_{\rm null}(\Theta) | \vec{d}_{\rm null}-\vec{g}_{\rm null}(\Theta)\rangle,
\end{aligned}
\label{eq:null_likelihood}
\end{equation}
where $\vec{g}_{\rm null}(\Theta)$ denotes the template for modeling the glitch $\vec{g}_{\rm null}$ 
given the model parameters $\Theta$. We then feed the null stream data to the glitch characterization tool explained in the previous section but 
without a signal model.\\
%We stress that the null stream based approach is independent of the glitch characterization tool used and can be adapted to any variable-dimensional sampling method. \\

\begin{figure}[h!]
    \centering
    \includegraphics[width=0.4\textwidth]{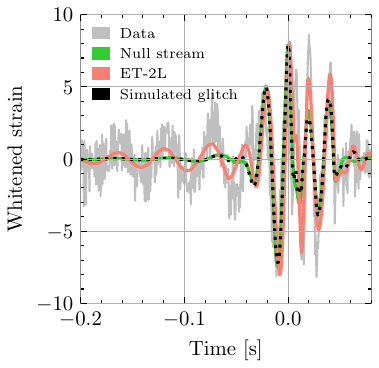}
    \caption{Green (red) indicates the glitch reconstructed using the \ETT~null stream (ET-2L) 
    for the $\Delta t = 0$ case. Grey shows the strain, and black shows the simulated glitch. The null stream helps avoid glitch mismodeling, resulting in much closer agreement with the simulated glitch.}
    \label{fig:overlap_reconstruction_panel}
\end{figure}

\begin{figure*}[t]
    \centering
    \includegraphics[width=\textwidth]{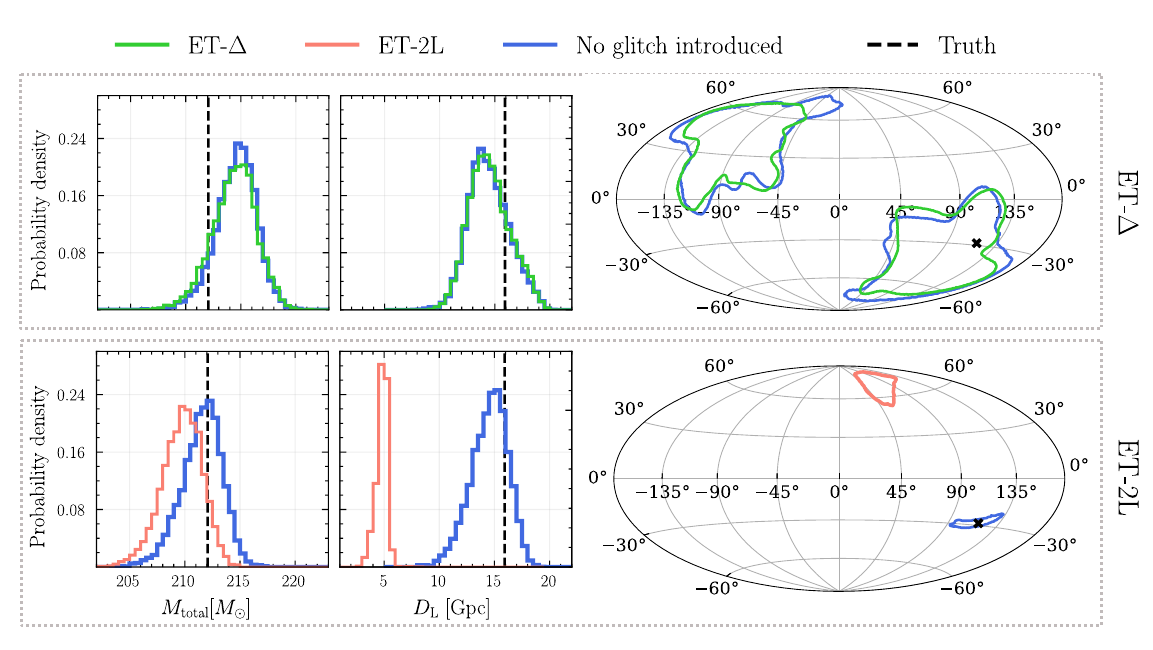}
  \caption{Measurement of detector-frame total mass $M_{\rm total}$, luminosity distance $D_{\rm L}$, 
  and sky localization. Green (red) color represent posterior distributions when the glitch is 
  removed using the \ETT~null stream (ET-2L). Blue color shows the measurements when no glitch is 
  introduced, serving as a benchmark for comparison, and dashed black lines indicate the true values. 
  The green posteriors accurately recover the true parameter values and show high consistency 
  with the benchmark. Though the red posterior for the total mass $M_{\rm total}$ recovers the true value, it shows clear signs of bias and widening. In contrast, the measurements of luminosity distance and sky location are biased and miss the true values.}
  \label{fig:corner_plot}
\end{figure*}

\section{Simulations} We simulate a binary black hole coalescence (BBH) with component masses  
$38\,M_\odot$ and $33\,M_\odot$ using the \textsc{IMRPhenomD} waveform 
model~\cite{Husa:2015iqa, Khan:2015jqa}. The source is placed at redshift of $\approx2$, near where the BBH merger 
rate is expected to peak~\cite{Belczynski:2016ieo, Oguri:2018muv}, making it a typical 
GW source. This setup yields a network 
signal-to-noise ratio (SNR) of 83 in \ETT~which is equally\footnote{The null stream is proportional to sum of antenna pattern functions whereas the SNR is proportional to the quadratic sum. 
Therefore, the individual detectors can have similar SNRs even though the sum of the signals in three 
detectors is zero.} distributed across ${\rm ET}_1$, ${\rm ET}_2$, ${\rm ET}_3$. Ref.~\cite{Branchesi:2023mws} estimates that $\mathcal{O}(10^4)$ such high-SNR events will be 
observed annually. The \ETT~is located at Virgo site with an arm length of 10 km. Gaussian noise is 
generated for all detectors using the ET-D sensitivity curve~\cite{Hild:2010id}. We analyze 4 s of data 
around the merger, with a lower frequency cut-off of 20 Hz and a sampling rate of 2048 Hz, employing 
standard signal parameter priors~\cite{Ashton:2018jfp, Romero-Shaw:2020owr}.

A blip glitch simulated using \texttt{gengli}~\cite{Lopez:2022lkd, Lopez:2022dho, Lopez:gengli_url} is then introduced into ${\rm ET}_1$. The glitch is scaled such that it has an SNR of $\sim 47$ which is similar to the SNR of the GW signal in ${\rm ET}_1$.

We generate 9 instances of data where the blip glitch overlaps with the GW signal. This consists of instance where the glitch onset conincides with the merger time of the signal as well as away from it. Across all instances, the signal and glitch morphologies, and noise realizations remain identical. The only varying 
quantity is the time separation $\Delta t$ between the merger time of the signal and the onset of the 
glitch.

We then perform glitch reconstruction using the null stream of \ETT.~A representative 
glitch template is built using the median posterior samples of the glitch model 
parameters.~This template is then subtracted from the ${\rm ET}_1$ data. Next, we perform parameter 
estimation using the cleaned ${\rm ET}_1$ data combined with the unaltered data from ${\rm ET}_2$ and 
${\rm ET}_3$ to measure 
the GW signal parameters using \texttt{Bilby}~\cite{Ashton:2018jfp, Romero-Shaw:2020owr}.\\

To compare the \ETT~and ET-2L configurations, we simulate a similar set-up for the latter 
configuration, perform the glitch subtraction, and measure GW signal parameters. The ET-2L configuration 
consists of two L-shaped misaligned interferometers with 15 km arm length located in Sardinia and the 
Euregion Meuse-Rhine (EMR) respectively. The GW signal parameters and the glitch morphology are the same as \ETT,~leading to a GW network SNR 
of 124, with the Sardinia and EMR interferometers contributing SNRs of 90 and 85, respectively. 
The glitch is introduced in the Sardinia interferometer, the SNR of which matches the glitch SNR in ${\rm ET}_1$.~In the absence of 
null stream in ET-2L, the glitch is reconstructed by simultaneously modeling the GW signal and 
the glitch assisted by the \textsc{IMRPhenomD} waveform model and using data from both interferometers.\\

\begin{figure}[t]
    \centering
    \includegraphics[width=0.45\textwidth]{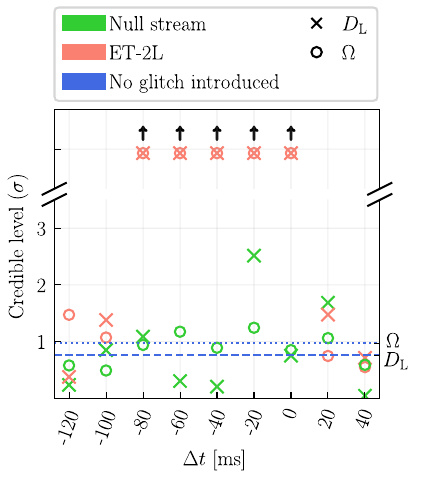}
    \caption{A comparison of the credible level of the true parameter value relative to the posterior 
    distributions for various time intervals $\Delta t$ between glitch onset and GW merger. 
    Credible levels are in standard deviation units $\sigma$, following a standard normal 
    distribution. The parameters considered are luminosity distance $D_{\rm L}$ and sky location 
    $\Omega$. Green (red) markers denote credible levels for the \ETT~null stream (ET-2L) based measurements and the blue lines provide credible levels in the absence of a 
    glitch as a benchmark (dotted for $\Omega$, dashed for $D_{\rm L}$). For the null stream based appraoch, the credible levels closely align with the benchmark, indicating accurate measurements. The posteriors for ET-2L lead to wider credible levels and generally exclude the true value, 
    except for the edge cases where the glitch is sufficiently distant in time from the GW merger.}
    \label{fig:trend}
\end{figure}

\section{Results} A visualization of the glitch reconstruction is presented in 
Fig.~\ref{fig:overlap_reconstruction_panel}. The \ETT~configuration, due to the null stream, achieves 
a more accurate glitch reconstruction by avoiding mismodeling from the GW signal. In contrast, 
for the ET-2L configuration inevitably some additional features from the signal are incurred. 

Fig.~\ref{fig:corner_plot} presents a comparison of the posterior distributions of the GW signal 
parameters after the glitch is mitigated, for 
$\Delta t = 0$. For all parameters shown, glitch mitigation with null stream for \ETT~results in posteriors 
that demonstrate high consistency and closely align with the benchmark results, 
where ``benchmark'' refers to a scenario when no glitch is introduced to the data.

For the ET-2L configuration, though the total mass $M_{\rm total}$ posterior shows clear signs of 
widening, it recovers the true value. This can be attributed to the SNR 
in the glitch-free interferometer, enabling 
intrinsic parameter recovery. In contrast, posteriors for extrinsic parameters, 
namely luminosity distance and sky location, are inconsistent with the benchmark's posteriors and miss the true value. 

Fig.~\ref{fig:trend} shows that the characteristics of extrinsic parameter recovery remain consistent 
across different $\Delta t$ values. A negative (positive) value of $\Delta t$ implies that the glitch 
onset falls before (after) the GW merger. For all time intervals considered, the~\ETT~configuration 
yields measurements where the true distance and sky location lie within $3\sigma$ and closely 
align with the no-glitch benchmark. In contrast, with the ET-2L configuration, the resulting measurements 
are wider or generally exclude the true values, except when the glitch onset is 100 ms before or 20 ms after the merger. 
%This means a glitch onset before the merger are more likely to bias the measurements. 

Besides providing the ability to remove glitches overlapping with the signals \textit{without} requiring any 
prior knowledge about the signal model, the null stream based approach of \ETT~also enhances 
the computational efficiency. It eliminates the need to simultaneously model both the glitch and 
the GW signal, and the latter would otherwise introduce an additional $\geq 11$ correlated parameters. 
The reduced dimensionality makes the null stream based approach approximately ten times faster than 
the alternate approach. Furthermore, the null stream ensures that the computational cost does not 
scale with the long-duration, loud, and overlapping signals, making it well-suited for the demands of 
the 3G era~\cite{Samajdar:2021egv, Pizzati:2021apa, Antonelli:2021vwg, Janquart:2022fzz, Relton:2021cax}.\\
%\section{Conclusions}

\section{Summary and discussion} In this work, we have demonstrated that the null stream, a feature inherent 
to Einstein Telescope's triangular configuration, enables a robust, third-generation-ready method for removing signal-overlapping glitches, which leads to parameter measurements that are essentially of the same quality as when no glitch is present. Apart from reconstructing glitches with 
high accuracy even when they are arbitrarily close in time with the peak of a GW signal, a substantial 
computational speed-up is achieved. While the loudness of the glitch with respect to the noise is reduced by a factor of $\simeq \sqrt{3}$ in the null 
stream, it does not 
affect parameter measurements shown in this work.

Without the null stream, the sky location and luminosity distance measurements of the GW source suffer from 
significant biases due to mismodeling of glitches even if the signal model is accurately known. 
Such biases could be harmful to science cases measuring cosmological parameters, in particular the ``dark siren'' method, which 
relies on three-dimensional volumetric matching between GW detections and galaxy 
catalogs~\cite{Schutz:1986gp,MacLeod:2007jd,DelPozzo:2011vcw,Chen:2017rfc,LIGOScientific:2018gmd,Gray:2019ksv,DES:2019ccw,DES:2020nay,LIGOScientific:2019zcs,Finke:2021aom,LIGOScientific:2021aug,Palmese:2021mjm}; 
the null stream will be of help here. Similarly, ET is anticipated to reconstruct the merger rate as a 
function of redshift (converting distance to redshift using a cosmological model) \cite{Branchesi:2023mws}; 
also this science case will benefit from the null stream.  Searches for anomalous dispersion of 
gravitational waves due to violations of general relativity represent another avenue that 
could be affected by incorrect distance 
measurements~\cite{LIGOScientific:2017bnn, LIGOScientific:2018dkp, LIGOScientific:2021sio, Narola:2023viz}. 
We note that in studies where information from multiple GW sources is combined, it will often be a 
relatively limited set of loudest signals that drive the combined result (tests of general relativity 
being a clear example \cite{LIGOScientific:2021sio}). If measurements on even a small fraction of these events are corrupted, it could adversely affect 
inference on the population as a whole. 	

Adding Cosmic Explorer to the network will significantly enhance sky localization and 
distance measurement due to the increased signal-to-noise ratio and trans-Atlantic 
baseline~\cite{Gupta:2023lga, Nitz:2021pbr}. However, we anticipate that the glitch mitigation capability of ET's null stream will still offer substantial improvement in parameter estimation compared to a network without null stream. 
For instance, having a null-stream-inherent ET would be particularly valuable for GW signals 
with considerable modeling uncertainties, such as core-collapse supernovae 
signals~\cite{Abdikamalov_2021}, signals from the post-merger phase of binary neutron 
stars~\cite{Dietrich:2020eud}, or signals influenced by microlensing, environmental factors, or 
orbital eccentricity~\cite{Roy:2024rhe, Leong:2023nuk, Wright:2021cbn, Gupte:2024jfe}. 
Such signals are of particular interest for the ET science case~\cite{Abac:2025saz}. A precise 
quantification of the advantages of a wider network with a null stream remains a subject for 
further investigation. In future, we also plan to explore the possibility of multiple detectors facing coincident glitches and strategies to mitigate such glitches. 
In this work, we have compared the triangular configuration of ET with the two distant, misaligned, L-shaped configuration in the context of glitch mitigation. One may add to the comparison the network of two distant but aligned L-shaped detectors, though it has been disfavoured in comparison to the misaligned configuration in the recent design study~\cite{Branchesi:2023mws}.

In summary, glitches that overlap with GW signals already pose a challenge for current detectors, 
and could significantly hinder the transition to an era of precision science. Robust glitch mitigation 
will be crucial for fully exploiting the potential of next-generation GW observatories in terms of 
astrophysics, fundamental physics, and cosmology. The null stream of a triangular Einstein Telescope 
may provide a key avenue towards this.\\

\begin{acknowledgments}
    We thank Stefano Schmidt, Mick Wright, Tomasz Baka, Soumen Roy, Sumit Kumar, Bhooshan Gadre, and Sophie Hourihane for discussions that led to improvement of this work. H.N., T.W., L.N., M.L., P.T.H.P, A.S., and C.V.D.B. are supported by the research programme of the Netherlands Organisation for Scientific Research~(NWO). This work is also partially supported by the Research Foundation - Flanders~(FWO) through Grant No. I002123N.~M.W. is supported by the FWO through Grant No.~11POK24N. This material is based upon work supported by NSF's LIGO Laboratory which is a major facility fully funded by the National Science Foundation. This research has made use of data, software and/or web tools obtained from the Gravitational Wave Open Science Center (https://www.gw-openscience.org), a service of LIGO Laboratory, the LIGO Scientific Collaboration and the Virgo Collaboration. LIGO is funded by the U.S. National Science Foundation. Virgo is funded by the French Centre National de Recherche Scientifique (CNRS), the Italian Istituto Nazionale della Fisica Nucleare (INFN) and the Dutch Nikhef, with contributions by Polish and Hungarian institutes.
\end{acknowledgments}

\bibliographystyle{apsrev4-1}
\bibliography{references}

\begin{thebibliography}{92}%
\makeatletter
\providecommand \@ifxundefined [1]{%
 \@ifx{#1\undefined}
}%
\providecommand \@ifnum [1]{%
 \ifnum #1\expandafter \@firstoftwo
 \else \expandafter \@secondoftwo
 \fi
}%
\providecommand \@ifx [1]{%
 \ifx #1\expandafter \@firstoftwo
 \else \expandafter \@secondoftwo
 \fi
}%
\providecommand \natexlab [1]{#1}%
\providecommand \enquote  [1]{``#1''}%
\providecommand \bibnamefont  [1]{#1}%
\providecommand \bibfnamefont [1]{#1}%
\providecommand \citenamefont [1]{#1}%
\providecommand \href@noop [0]{\@secondoftwo}%
\providecommand \href [0]{\begingroup \@sanitize@url \@href}%
\providecommand \@href[1]{\@@startlink{#1}\@@href}%
\providecommand \@@href[1]{\endgroup#1\@@endlink}%
\providecommand \@sanitize@url [0]{\catcode `\\12\catcode `\$12\catcode
  `\&12\catcode `\#12\catcode `\^12\catcode `\_12\catcode `\%12\relax}%
\providecommand \@@startlink[1]{}%
\providecommand \@@endlink[0]{}%
\providecommand \url  [0]{\begingroup\@sanitize@url \@url }%
\providecommand \@url [1]{\endgroup\@href {#1}{\urlprefix }}%
\providecommand \urlprefix  [0]{URL }%
\providecommand \Eprint [0]{\href }%
\providecommand \doibase [0]{http://dx.doi.org/}%
\providecommand \selectlanguage [0]{\@gobble}%
\providecommand \bibinfo  [0]{\@secondoftwo}%
\providecommand \bibfield  [0]{\@secondoftwo}%
\providecommand \translation [1]{[#1]}%
\providecommand \BibitemOpen [0]{}%
\providecommand \bibitemStop [0]{}%
\providecommand \bibitemNoStop [0]{.\EOS\space}%
\providecommand \EOS [0]{\spacefactor3000\relax}%
\providecommand \BibitemShut  [1]{\csname bibitem#1\endcsname}%
\let\auto@bib@innerbib\@empty
%</preamble>
\bibitem [{\citenamefont {Aasi}\ \emph {et~al.}(2015)\citenamefont {Aasi} \emph
  {et~al.}}]{LIGOScientific:2014pky}%
  \BibitemOpen
  \bibfield  {author} {\bibinfo {author} {\bibfnamefont {J.}~\bibnamefont
  {Aasi}} \emph {et~al.} (\bibinfo {collaboration} {LIGO Scientific}),\ }\href
  {\doibase 10.1088/0264-9381/32/7/074001} {\bibfield  {journal} {\bibinfo
  {journal} {Class. Quant. Grav.}\ }\textbf {\bibinfo {volume} {32}},\ \bibinfo
  {pages} {074001} (\bibinfo {year} {2015})},\ \Eprint
  {http://arxiv.org/abs/1411.4547} {arXiv:1411.4547 [gr-qc]} \BibitemShut
  {NoStop}%
\bibitem [{\citenamefont {Acernese}\ \emph {et~al.}(2015)\citenamefont
  {Acernese} \emph {et~al.}}]{VIRGO:2014yos}%
  \BibitemOpen
  \bibfield  {author} {\bibinfo {author} {\bibfnamefont {F.}~\bibnamefont
  {Acernese}} \emph {et~al.} (\bibinfo {collaboration} {VIRGO}),\ }\href
  {\doibase 10.1088/0264-9381/32/2/024001} {\bibfield  {journal} {\bibinfo
  {journal} {Class. Quant. Grav.}\ }\textbf {\bibinfo {volume} {32}},\ \bibinfo
  {pages} {024001} (\bibinfo {year} {2015})},\ \Eprint
  {http://arxiv.org/abs/1408.3978} {arXiv:1408.3978 [gr-qc]} \BibitemShut
  {NoStop}%
\bibitem [{\citenamefont {Abbott}\ \emph
  {et~al.}(2021{\natexlab{a}})\citenamefont {Abbott} \emph
  {et~al.}}]{LIGOScientific:2021aug}%
  \BibitemOpen
  \bibfield  {author} {\bibinfo {author} {\bibfnamefont {R.}~\bibnamefont
  {Abbott}} \emph {et~al.} (\bibinfo {collaboration} {LIGO Scientific, VIRGO,
  KAGRA}),\ }\href@noop {} {\  (\bibinfo {year} {2021}{\natexlab{a}})},\
  \Eprint {http://arxiv.org/abs/2111.03604} {arXiv:2111.03604 [astro-ph.CO]}
  \BibitemShut {NoStop}%
\bibitem [{\citenamefont {Abbott}\ \emph
  {et~al.}(2023{\natexlab{a}})\citenamefont {Abbott} \emph
  {et~al.}}]{KAGRA:2021duu}%
  \BibitemOpen
  \bibfield  {author} {\bibinfo {author} {\bibfnamefont {R.}~\bibnamefont
  {Abbott}} \emph {et~al.} (\bibinfo {collaboration} {KAGRA, VIRGO, LIGO
  Scientific}),\ }\href {\doibase 10.1103/PhysRevX.13.011048} {\bibfield
  {journal} {\bibinfo  {journal} {Phys. Rev. X}\ }\textbf {\bibinfo {volume}
  {13}},\ \bibinfo {pages} {011048} (\bibinfo {year} {2023}{\natexlab{a}})},\
  \Eprint {http://arxiv.org/abs/2111.03634} {arXiv:2111.03634 [astro-ph.HE]}
  \BibitemShut {NoStop}%
\bibitem [{\citenamefont {Abbott}\ \emph
  {et~al.}(2021{\natexlab{b}})\citenamefont {Abbott} \emph
  {et~al.}}]{LIGOScientific:2021sio}%
  \BibitemOpen
  \bibfield  {author} {\bibinfo {author} {\bibfnamefont {R.}~\bibnamefont
  {Abbott}} \emph {et~al.} (\bibinfo {collaboration} {LIGO Scientific, VIRGO,
  KAGRA}),\ }\href@noop {} {\  (\bibinfo {year} {2021}{\natexlab{b}})},\
  \Eprint {http://arxiv.org/abs/2112.06861} {arXiv:2112.06861 [gr-qc]}
  \BibitemShut {NoStop}%
\bibitem [{\citenamefont {Abbott}\ \emph
  {et~al.}(2023{\natexlab{b}})\citenamefont {Abbott} \emph
  {et~al.}}]{LIGOScientific:2023bwz}%
  \BibitemOpen
  \bibfield  {author} {\bibinfo {author} {\bibfnamefont {R.}~\bibnamefont
  {Abbott}} \emph {et~al.} (\bibinfo {collaboration} {LIGO Scientific, VIRGO,
  KAGRA}),\ }\href@noop {} {\  (\bibinfo {year} {2023}{\natexlab{b}})},\
  \Eprint {http://arxiv.org/abs/2304.08393} {arXiv:2304.08393 [gr-qc]}
  \BibitemShut {NoStop}%
\bibitem [{\citenamefont {Ashton}\ \emph {et~al.}(2019)\citenamefont {Ashton}
  \emph {et~al.}}]{Ashton:2018jfp}%
  \BibitemOpen
  \bibfield  {author} {\bibinfo {author} {\bibfnamefont {G.}~\bibnamefont
  {Ashton}} \emph {et~al.},\ }\href {\doibase 10.3847/1538-4365/ab06fc}
  {\bibfield  {journal} {\bibinfo  {journal} {Astrophys. J. Suppl.}\ }\textbf
  {\bibinfo {volume} {241}},\ \bibinfo {pages} {27} (\bibinfo {year} {2019})},\
  \Eprint {http://arxiv.org/abs/1811.02042} {arXiv:1811.02042 [astro-ph.IM]}
  \BibitemShut {NoStop}%
\bibitem [{\citenamefont {Romero-Shaw}\ \emph {et~al.}(2020)\citenamefont
  {Romero-Shaw} \emph {et~al.}}]{Romero-Shaw:2020owr}%
  \BibitemOpen
  \bibfield  {author} {\bibinfo {author} {\bibfnamefont {I.~M.}\ \bibnamefont
  {Romero-Shaw}} \emph {et~al.},\ }\href {\doibase 10.1093/mnras/staa2850}
  {\bibfield  {journal} {\bibinfo  {journal} {Mon. Not. Roy. Astron. Soc.}\
  }\textbf {\bibinfo {volume} {499}},\ \bibinfo {pages} {3295} (\bibinfo {year}
  {2020})},\ \Eprint {http://arxiv.org/abs/2006.00714} {arXiv:2006.00714
  [astro-ph.IM]} \BibitemShut {NoStop}%
\bibitem [{\citenamefont {Narola}\ \emph {et~al.}(2023)\citenamefont {Narola}
  \emph {et~al.}}]{Narola:2023men}%
  \BibitemOpen
  \bibfield  {author} {\bibinfo {author} {\bibfnamefont {H.}~\bibnamefont
  {Narola}} \emph {et~al.},\ }\href@noop {} {\  (\bibinfo {year} {2023})},\
  \Eprint {http://arxiv.org/abs/2308.12140} {arXiv:2308.12140 [gr-qc]}
  \BibitemShut {NoStop}%
\bibitem [{\citenamefont {Abbott}\ \emph
  {et~al.}(2020{\natexlab{a}})\citenamefont {Abbott} \emph
  {et~al.}}]{LIGOScientific:2019hgc}%
  \BibitemOpen
  \bibfield  {author} {\bibinfo {author} {\bibfnamefont {B.~P.}\ \bibnamefont
  {Abbott}} \emph {et~al.} (\bibinfo {collaboration} {LIGO Scientific,
  Virgo}),\ }\href {\doibase 10.1088/1361-6382/ab685e} {\bibfield  {journal}
  {\bibinfo  {journal} {Class. Quant. Grav.}\ }\textbf {\bibinfo {volume}
  {37}},\ \bibinfo {pages} {055002} (\bibinfo {year} {2020}{\natexlab{a}})},\
  \Eprint {http://arxiv.org/abs/1908.11170} {arXiv:1908.11170 [gr-qc]}
  \BibitemShut {NoStop}%
\bibitem [{\citenamefont {Abbott}\ \emph {et~al.}(2018)\citenamefont {Abbott}
  \emph {et~al.}}]{LIGOScientific:2017tza}%
  \BibitemOpen
  \bibfield  {author} {\bibinfo {author} {\bibfnamefont {B.~P.}\ \bibnamefont
  {Abbott}} \emph {et~al.} (\bibinfo {collaboration} {LIGO Scientific,
  Virgo}),\ }\href {\doibase 10.1088/1361-6382/aaaafa} {\bibfield  {journal}
  {\bibinfo  {journal} {Class. Quant. Grav.}\ }\textbf {\bibinfo {volume}
  {35}},\ \bibinfo {pages} {065010} (\bibinfo {year} {2018})},\ \Eprint
  {http://arxiv.org/abs/1710.02185} {arXiv:1710.02185 [gr-qc]} \BibitemShut
  {NoStop}%
\bibitem [{\citenamefont {Davis}\ \emph {et~al.}(2021)\citenamefont {Davis}
  \emph {et~al.}}]{LIGO:2021ppb}%
  \BibitemOpen
  \bibfield  {author} {\bibinfo {author} {\bibfnamefont {D.}~\bibnamefont
  {Davis}} \emph {et~al.} (\bibinfo {collaboration} {LIGO}),\ }\href {\doibase
  10.1088/1361-6382/abfd85} {\bibfield  {journal} {\bibinfo  {journal} {Class.
  Quant. Grav.}\ }\textbf {\bibinfo {volume} {38}},\ \bibinfo {pages} {135014}
  (\bibinfo {year} {2021})},\ \Eprint {http://arxiv.org/abs/2101.11673}
  {arXiv:2101.11673 [astro-ph.IM]} \BibitemShut {NoStop}%
\bibitem [{\citenamefont {Acernese}\ \emph {et~al.}(2022)\citenamefont
  {Acernese} \emph {et~al.}}]{Virgo:2022fxr}%
  \BibitemOpen
  \bibfield  {author} {\bibinfo {author} {\bibfnamefont {F.}~\bibnamefont
  {Acernese}} \emph {et~al.} (\bibinfo {collaboration} {Virgo}),\ }\href@noop
  {} {\  (\bibinfo {year} {2022})},\ \Eprint {http://arxiv.org/abs/2205.01555}
  {arXiv:2205.01555 [gr-qc]} \BibitemShut {NoStop}%
\bibitem [{\citenamefont {Soni}\ \emph {et~al.}(2024)\citenamefont {Soni} \emph
  {et~al.}}]{LIGO:2024kkz}%
  \BibitemOpen
  \bibfield  {author} {\bibinfo {author} {\bibfnamefont {S.}~\bibnamefont
  {Soni}} \emph {et~al.} (\bibinfo {collaboration} {LIGO}),\ }\href@noop {} {\
  (\bibinfo {year} {2024})},\ \Eprint {http://arxiv.org/abs/2409.02831}
  {arXiv:2409.02831 [astro-ph.IM]} \BibitemShut {NoStop}%
\bibitem [{\citenamefont {{Soni}}\ and\ \citenamefont {others {LIGO Scientific
  Collaboration}}(2021)}]{LIGO:2020zwl}%
  \BibitemOpen
  \bibfield  {author} {\bibinfo {author} {\bibfnamefont {S.}~\bibnamefont
  {{Soni}}}\ and\ \bibinfo {author} {\bibnamefont {others {LIGO Scientific
  Collaboration}}},\ }\href {\doibase 10.1088/1361-6382/abc906} {\bibfield
  {journal} {\bibinfo  {journal} {Classical and Quantum Gravity}\ }\textbf
  {\bibinfo {volume} {38}},\ \bibinfo {eid} {025016} (\bibinfo {year}
  {2021})},\ \Eprint {http://arxiv.org/abs/2007.14876} {arXiv:2007.14876
  [astro-ph.IM]} \BibitemShut {NoStop}%
\bibitem [{\citenamefont {Cabero}\ \emph {et~al.}(2019)\citenamefont {Cabero}
  \emph {et~al.}}]{Cabero:2019orq}%
  \BibitemOpen
  \bibfield  {author} {\bibinfo {author} {\bibfnamefont {M.}~\bibnamefont
  {Cabero}} \emph {et~al.},\ }\href {\doibase 10.1088/1361-6382/ab2e14}
  {\bibfield  {journal} {\bibinfo  {journal} {Class. Quant. Grav.}\ }\textbf
  {\bibinfo {volume} {36}},\ \bibinfo {pages} {15} (\bibinfo {year} {2019})},\
  \Eprint {http://arxiv.org/abs/1901.05093} {arXiv:1901.05093
  [physics.ins-det]} \BibitemShut {NoStop}%
\bibitem [{\citenamefont {Pankow}\ \emph {et~al.}(2018)\citenamefont {Pankow}
  \emph {et~al.}}]{Pankow:2018qpo}%
  \BibitemOpen
  \bibfield  {author} {\bibinfo {author} {\bibfnamefont {C.}~\bibnamefont
  {Pankow}} \emph {et~al.},\ }\href {\doibase 10.1103/PhysRevD.98.084016}
  {\bibfield  {journal} {\bibinfo  {journal} {Phys. Rev. D}\ }\textbf {\bibinfo
  {volume} {98}},\ \bibinfo {pages} {084016} (\bibinfo {year} {2018})},\
  \Eprint {http://arxiv.org/abs/1808.03619} {arXiv:1808.03619 [gr-qc]}
  \BibitemShut {NoStop}%
\bibitem [{\citenamefont {Payne}\ \emph {et~al.}(2022)\citenamefont {Payne}
  \emph {et~al.}}]{Payne:2022spz}%
  \BibitemOpen
  \bibfield  {author} {\bibinfo {author} {\bibfnamefont {E.}~\bibnamefont
  {Payne}} \emph {et~al.},\ }\href {\doibase 10.1103/PhysRevD.106.104017}
  {\bibfield  {journal} {\bibinfo  {journal} {Phys. Rev. D}\ }\textbf {\bibinfo
  {volume} {106}},\ \bibinfo {pages} {104017} (\bibinfo {year} {2022})},\
  \Eprint {http://arxiv.org/abs/2206.11932} {arXiv:2206.11932 [gr-qc]}
  \BibitemShut {NoStop}%
\bibitem [{\citenamefont {Udall}\ \emph {et~al.}(2024)\citenamefont {Udall}
  \emph {et~al.}}]{Udall:2024ovp}%
  \BibitemOpen
  \bibfield  {author} {\bibinfo {author} {\bibfnamefont {R.}~\bibnamefont
  {Udall}} \emph {et~al.},\ }\href@noop {} {\  (\bibinfo {year} {2024})},\
  \Eprint {http://arxiv.org/abs/2409.03912} {arXiv:2409.03912 [gr-qc]}
  \BibitemShut {NoStop}%
\bibitem [{\citenamefont {Gupta}\ \emph
  {et~al.}(2024{\natexlab{a}})\citenamefont {Gupta} \emph
  {et~al.}}]{Gupta:2024gun}%
  \BibitemOpen
  \bibfield  {author} {\bibinfo {author} {\bibfnamefont {A.}~\bibnamefont
  {Gupta}} \emph {et~al.},\ }\href@noop {} {\  (\bibinfo {year}
  {2024}{\natexlab{a}})},\ \Eprint {http://arxiv.org/abs/2405.02197}
  {arXiv:2405.02197 [gr-qc]} \BibitemShut {NoStop}%
\bibitem [{\citenamefont {Abbott}\ \emph
  {et~al.}(2023{\natexlab{c}})\citenamefont {Abbott} \emph
  {et~al.}}]{KAGRA:2021vkt}%
  \BibitemOpen
  \bibfield  {author} {\bibinfo {author} {\bibfnamefont {R.}~\bibnamefont
  {Abbott}} \emph {et~al.} (\bibinfo {collaboration} {KAGRA, VIRGO, LIGO
  Scientific}),\ }\href {\doibase 10.1103/PhysRevX.13.041039} {\bibfield
  {journal} {\bibinfo  {journal} {Phys. Rev. X}\ }\textbf {\bibinfo {volume}
  {13}},\ \bibinfo {pages} {041039} (\bibinfo {year} {2023}{\natexlab{c}})},\
  \Eprint {http://arxiv.org/abs/2111.03606} {arXiv:2111.03606 [gr-qc]}
  \BibitemShut {NoStop}%
\bibitem [{\citenamefont {Abbott}\ \emph
  {et~al.}(2019{\natexlab{a}})\citenamefont {Abbott} \emph
  {et~al.}}]{LIGOScientific:2018mvr}%
  \BibitemOpen
  \bibfield  {author} {\bibinfo {author} {\bibfnamefont {B.~P.}\ \bibnamefont
  {Abbott}} \emph {et~al.} (\bibinfo {collaboration} {LIGO Scientific,
  Virgo}),\ }\href {\doibase 10.1103/PhysRevX.9.031040} {\bibfield  {journal}
  {\bibinfo  {journal} {Phys. Rev. X}\ }\textbf {\bibinfo {volume} {9}},\
  \bibinfo {pages} {031040} (\bibinfo {year} {2019}{\natexlab{a}})},\ \Eprint
  {http://arxiv.org/abs/1811.12907} {arXiv:1811.12907 [astro-ph.HE]}
  \BibitemShut {NoStop}%
\bibitem [{\citenamefont {Abbott}\ \emph
  {et~al.}(2021{\natexlab{c}})\citenamefont {Abbott} \emph
  {et~al.}}]{LIGOScientific:2020ibl}%
  \BibitemOpen
  \bibfield  {author} {\bibinfo {author} {\bibfnamefont {R.}~\bibnamefont
  {Abbott}} \emph {et~al.} (\bibinfo {collaboration} {LIGO Scientific,
  Virgo}),\ }\href {\doibase 10.1103/PhysRevX.11.021053} {\bibfield  {journal}
  {\bibinfo  {journal} {Phys. Rev. X}\ }\textbf {\bibinfo {volume} {11}},\
  \bibinfo {pages} {021053} (\bibinfo {year} {2021}{\natexlab{c}})},\ \Eprint
  {http://arxiv.org/abs/2010.14527} {arXiv:2010.14527 [gr-qc]} \BibitemShut
  {NoStop}%
\bibitem [{\citenamefont {Abbott}\ \emph
  {et~al.}(2017{\natexlab{a}})\citenamefont {Abbott} \emph
  {et~al.}}]{LIGOScientific:2017vwq}%
  \BibitemOpen
  \bibfield  {author} {\bibinfo {author} {\bibfnamefont {B.~P.}\ \bibnamefont
  {Abbott}} \emph {et~al.} (\bibinfo {collaboration} {LIGO Scientific,
  Virgo}),\ }\href {\doibase 10.1103/PhysRevLett.119.161101} {\bibfield
  {journal} {\bibinfo  {journal} {Phys. Rev. Lett.}\ }\textbf {\bibinfo
  {volume} {119}},\ \bibinfo {pages} {161101} (\bibinfo {year}
  {2017}{\natexlab{a}})},\ \Eprint {http://arxiv.org/abs/1710.05832}
  {arXiv:1710.05832 [gr-qc]} \BibitemShut {NoStop}%
\bibitem [{\citenamefont {Chatziioannou}\ \emph {et~al.}(2021)\citenamefont
  {Chatziioannou} \emph {et~al.}}]{Chatziioannou:2021ezd}%
  \BibitemOpen
  \bibfield  {author} {\bibinfo {author} {\bibfnamefont {K.}~\bibnamefont
  {Chatziioannou}} \emph {et~al.},\ }\href {\doibase
  10.1103/PhysRevD.103.044013} {\bibfield  {journal} {\bibinfo  {journal}
  {Phys. Rev. D}\ }\textbf {\bibinfo {volume} {103}},\ \bibinfo {pages}
  {044013} (\bibinfo {year} {2021})},\ \Eprint
  {http://arxiv.org/abs/2101.01200} {arXiv:2101.01200 [gr-qc]} \BibitemShut
  {NoStop}%
\bibitem [{\citenamefont {Punturo}\ \emph {et~al.}(2010)\citenamefont {Punturo}
  \emph {et~al.}}]{Punturo:2010zz}%
  \BibitemOpen
  \bibfield  {author} {\bibinfo {author} {\bibfnamefont {M.}~\bibnamefont
  {Punturo}} \emph {et~al.},\ }\href {\doibase 10.1088/0264-9381/27/19/194002}
  {\bibfield  {journal} {\bibinfo  {journal} {Class. Quant. Grav.}\ }\textbf
  {\bibinfo {volume} {27}},\ \bibinfo {pages} {194002} (\bibinfo {year}
  {2010})}\BibitemShut {NoStop}%
\bibitem [{\citenamefont {Evans}\ \emph {et~al.}(2021)\citenamefont {Evans}
  \emph {et~al.}}]{Evans:2021gyd}%
  \BibitemOpen
  \bibfield  {author} {\bibinfo {author} {\bibfnamefont {M.}~\bibnamefont
  {Evans}} \emph {et~al.},\ }\href@noop {} {\  (\bibinfo {year} {2021})},\
  \Eprint {http://arxiv.org/abs/2109.09882} {arXiv:2109.09882 [astro-ph.IM]}
  \BibitemShut {NoStop}%
\bibitem [{\citenamefont {Hild}\ \emph {et~al.}(2011)\citenamefont {Hild} \emph
  {et~al.}}]{Hild:2010id}%
  \BibitemOpen
  \bibfield  {author} {\bibinfo {author} {\bibfnamefont {S.}~\bibnamefont
  {Hild}} \emph {et~al.},\ }\href {\doibase 10.1088/0264-9381/28/9/094013}
  {\bibfield  {journal} {\bibinfo  {journal} {Class. Quant. Grav.}\ }\textbf
  {\bibinfo {volume} {28}},\ \bibinfo {pages} {094013} (\bibinfo {year}
  {2011})},\ \Eprint {http://arxiv.org/abs/1012.0908} {arXiv:1012.0908 [gr-qc]}
  \BibitemShut {NoStop}%
\bibitem [{\citenamefont {Maggiore}\ \emph {et~al.}(2020)\citenamefont
  {Maggiore} \emph {et~al.}}]{Maggiore:2019uih}%
  \BibitemOpen
  \bibfield  {author} {\bibinfo {author} {\bibfnamefont {M.}~\bibnamefont
  {Maggiore}} \emph {et~al.},\ }\href {\doibase 10.1088/1475-7516/2020/03/050}
  {\bibfield  {journal} {\bibinfo  {journal} {JCAP}\ }\textbf {\bibinfo
  {volume} {03}},\ \bibinfo {pages} {050} (\bibinfo {year} {2020})},\ \Eprint
  {http://arxiv.org/abs/1912.02622} {arXiv:1912.02622 [astro-ph.CO]}
  \BibitemShut {NoStop}%
\bibitem [{\citenamefont {Reitze}\ \emph {et~al.}(2019)\citenamefont {Reitze}
  \emph {et~al.}}]{Reitze:2019iox}%
  \BibitemOpen
  \bibfield  {author} {\bibinfo {author} {\bibfnamefont {D.}~\bibnamefont
  {Reitze}} \emph {et~al.},\ }\href@noop {} {\bibfield  {journal} {\bibinfo
  {journal} {Bull. Am. Astron. Soc.}\ }\textbf {\bibinfo {volume} {51}},\
  \bibinfo {pages} {035} (\bibinfo {year} {2019})},\ \Eprint
  {http://arxiv.org/abs/1907.04833} {arXiv:1907.04833 [astro-ph.IM]}
  \BibitemShut {NoStop}%
\bibitem [{\citenamefont {Branchesi}\ \emph {et~al.}(2023)\citenamefont
  {Branchesi} \emph {et~al.}}]{Branchesi:2023mws}%
  \BibitemOpen
  \bibfield  {author} {\bibinfo {author} {\bibfnamefont {M.}~\bibnamefont
  {Branchesi}} \emph {et~al.},\ }\href@noop {} {\  (\bibinfo {year} {2023})},\
  \Eprint {http://arxiv.org/abs/2303.15923} {arXiv:2303.15923 [gr-qc]}
  \BibitemShut {NoStop}%
\bibitem [{\citenamefont {Aasi}\ \emph {et~al.}(2012)\citenamefont {Aasi} \emph
  {et~al.}}]{VIRGO:2012oxz}%
  \BibitemOpen
  \bibfield  {author} {\bibinfo {author} {\bibfnamefont {J.}~\bibnamefont
  {Aasi}} \emph {et~al.} (\bibinfo {collaboration} {VIRGO}),\ }\href {\doibase
  10.1088/0264-9381/29/15/155002} {\bibfield  {journal} {\bibinfo  {journal}
  {Class. Quant. Grav.}\ }\textbf {\bibinfo {volume} {29}},\ \bibinfo {pages}
  {155002} (\bibinfo {year} {2012})},\ \Eprint {http://arxiv.org/abs/1203.5613}
  {arXiv:1203.5613 [gr-qc]} \BibitemShut {NoStop}%
\bibitem [{\citenamefont {Nuttall}(2018)}]{Nuttall:2018xhi}%
  \BibitemOpen
  \bibfield  {author} {\bibinfo {author} {\bibfnamefont {L.~K.}\ \bibnamefont
  {Nuttall}},\ }\href {\doibase 10.1098/rsta.2017.0286} {\bibfield  {journal}
  {\bibinfo  {journal} {Phil. Trans. Roy. Soc. Lond. A}\ }\textbf {\bibinfo
  {volume} {376}},\ \bibinfo {pages} {20170286} (\bibinfo {year} {2018})},\
  \Eprint {http://arxiv.org/abs/1804.07592} {arXiv:1804.07592 [astro-ph.IM]}
  \BibitemShut {NoStop}%
\bibitem [{\citenamefont {Davis}\ and\ \citenamefont
  {Walker}(2022)}]{Davis:2022dnd}%
  \BibitemOpen
  \bibfield  {author} {\bibinfo {author} {\bibfnamefont {D.}~\bibnamefont
  {Davis}}\ and\ \bibinfo {author} {\bibfnamefont {M.}~\bibnamefont {Walker}},\
  }\href {\doibase 10.3390/galaxies10010012} {\bibfield  {journal} {\bibinfo
  {journal} {Galaxies}\ }\textbf {\bibinfo {volume} {10}},\ \bibinfo {pages}
  {12} (\bibinfo {year} {2022})}\BibitemShut {NoStop}%
\bibitem [{\citenamefont {Hourihane}\ \emph {et~al.}(2022)\citenamefont
  {Hourihane} \emph {et~al.}}]{Hourihane:2022doe}%
  \BibitemOpen
  \bibfield  {author} {\bibinfo {author} {\bibfnamefont {S.}~\bibnamefont
  {Hourihane}} \emph {et~al.},\ }\href {\doibase 10.1103/PhysRevD.106.042006}
  {\bibfield  {journal} {\bibinfo  {journal} {Phys. Rev. D}\ }\textbf {\bibinfo
  {volume} {106}},\ \bibinfo {pages} {042006} (\bibinfo {year} {2022})},\
  \Eprint {http://arxiv.org/abs/2205.13580} {arXiv:2205.13580 [gr-qc]}
  \BibitemShut {NoStop}%
\bibitem [{\citenamefont {Ghonge}\ \emph {et~al.}(2023)\citenamefont {Ghonge}
  \emph {et~al.}}]{Ghonge:2023ksb}%
  \BibitemOpen
  \bibfield  {author} {\bibinfo {author} {\bibfnamefont {S.}~\bibnamefont
  {Ghonge}} \emph {et~al.},\ }\href@noop {} {\  (\bibinfo {year} {2023})},\
  \Eprint {http://arxiv.org/abs/2311.09159} {arXiv:2311.09159 [gr-qc]}
  \BibitemShut {NoStop}%
\bibitem [{\citenamefont {Buonanno}\ and\ \citenamefont
  {Damour}(1999)}]{Buonanno:1998gg}%
  \BibitemOpen
  \bibfield  {author} {\bibinfo {author} {\bibfnamefont {A.}~\bibnamefont
  {Buonanno}}\ and\ \bibinfo {author} {\bibfnamefont {T.}~\bibnamefont
  {Damour}},\ }\href {\doibase 10.1103/PhysRevD.59.084006} {\bibfield
  {journal} {\bibinfo  {journal} {Phys. Rev. D}\ }\textbf {\bibinfo {volume}
  {59}},\ \bibinfo {pages} {084006} (\bibinfo {year} {1999})},\ \Eprint
  {http://arxiv.org/abs/gr-qc/9811091} {arXiv:gr-qc/9811091} \BibitemShut
  {NoStop}%
\bibitem [{\citenamefont {Husa}\ \emph {et~al.}(2016)\citenamefont {Husa} \emph
  {et~al.}}]{Husa:2015iqa}%
  \BibitemOpen
  \bibfield  {author} {\bibinfo {author} {\bibfnamefont {S.}~\bibnamefont
  {Husa}} \emph {et~al.},\ }\href {\doibase 10.1103/PhysRevD.93.044006}
  {\bibfield  {journal} {\bibinfo  {journal} {Phys. Rev. D}\ }\textbf {\bibinfo
  {volume} {93}},\ \bibinfo {pages} {044006} (\bibinfo {year} {2016})},\
  \Eprint {http://arxiv.org/abs/1508.07250} {arXiv:1508.07250 [gr-qc]}
  \BibitemShut {NoStop}%
\bibitem [{\citenamefont {Boyle}\ \emph {et~al.}(2019)\citenamefont {Boyle}
  \emph {et~al.}}]{Boyle:2019kee}%
  \BibitemOpen
  \bibfield  {author} {\bibinfo {author} {\bibfnamefont {M.}~\bibnamefont
  {Boyle}} \emph {et~al.},\ }\href {\doibase 10.1088/1361-6382/ab34e2}
  {\bibfield  {journal} {\bibinfo  {journal} {Class. Quant. Grav.}\ }\textbf
  {\bibinfo {volume} {36}},\ \bibinfo {pages} {195006} (\bibinfo {year}
  {2019})},\ \Eprint {http://arxiv.org/abs/1904.04831} {arXiv:1904.04831
  [gr-qc]} \BibitemShut {NoStop}%
\bibitem [{\citenamefont {Lopez}\ \emph {et~al.}(2024)\citenamefont {Lopez},
  \citenamefont {Caneva}, \citenamefont {Martins}, \citenamefont {Schmidt},
  \citenamefont {Schoppink}, \citenamefont {van Straalen}, \citenamefont
  {Capano},\ and\ \citenamefont {Caudill}}]{Lopez:2024xby}%
  \BibitemOpen
  \bibfield  {author} {\bibinfo {author} {\bibfnamefont {M.}~\bibnamefont
  {Lopez}}, \bibinfo {author} {\bibfnamefont {G.}~\bibnamefont {Caneva}},
  \bibinfo {author} {\bibfnamefont {A.}~\bibnamefont {Martins}}, \bibinfo
  {author} {\bibfnamefont {S.}~\bibnamefont {Schmidt}}, \bibinfo {author}
  {\bibfnamefont {J.}~\bibnamefont {Schoppink}}, \bibinfo {author}
  {\bibfnamefont {W.}~\bibnamefont {van Straalen}}, \bibinfo {author}
  {\bibfnamefont {C.}~\bibnamefont {Capano}}, \ and\ \bibinfo {author}
  {\bibfnamefont {S.}~\bibnamefont {Caudill}},\ }\href@noop {} {\  (\bibinfo
  {year} {2024})},\ \Eprint {http://arxiv.org/abs/2412.17169} {arXiv:2412.17169
  [astro-ph.IM]} \BibitemShut {NoStop}%
\bibitem [{\citenamefont {Samajdar}\ \emph {et~al.}(2021)\citenamefont
  {Samajdar} \emph {et~al.}}]{Samajdar:2021egv}%
  \BibitemOpen
  \bibfield  {author} {\bibinfo {author} {\bibfnamefont {A.}~\bibnamefont
  {Samajdar}} \emph {et~al.},\ }\href {\doibase 10.1103/PhysRevD.104.044003}
  {\bibfield  {journal} {\bibinfo  {journal} {Phys. Rev. D}\ }\textbf {\bibinfo
  {volume} {104}},\ \bibinfo {pages} {044003} (\bibinfo {year} {2021})},\
  \Eprint {http://arxiv.org/abs/2102.07544} {arXiv:2102.07544 [gr-qc]}
  \BibitemShut {NoStop}%
\bibitem [{\citenamefont {Pizzati}\ \emph {et~al.}(2022)\citenamefont {Pizzati}
  \emph {et~al.}}]{Pizzati:2021apa}%
  \BibitemOpen
  \bibfield  {author} {\bibinfo {author} {\bibfnamefont {E.}~\bibnamefont
  {Pizzati}} \emph {et~al.},\ }\href {\doibase 10.1103/PhysRevD.105.104016}
  {\bibfield  {journal} {\bibinfo  {journal} {Phys. Rev. D}\ }\textbf {\bibinfo
  {volume} {105}},\ \bibinfo {pages} {104016} (\bibinfo {year} {2022})},\
  \Eprint {http://arxiv.org/abs/2102.07692} {arXiv:2102.07692 [gr-qc]}
  \BibitemShut {NoStop}%
\bibitem [{\citenamefont {Antonelli}\ \emph {et~al.}(2021)\citenamefont
  {Antonelli}, \citenamefont {Burke},\ and\ \citenamefont
  {Gair}}]{Antonelli:2021vwg}%
  \BibitemOpen
  \bibfield  {author} {\bibinfo {author} {\bibfnamefont {A.}~\bibnamefont
  {Antonelli}}, \bibinfo {author} {\bibfnamefont {O.}~\bibnamefont {Burke}}, \
  and\ \bibinfo {author} {\bibfnamefont {J.~R.}\ \bibnamefont {Gair}},\ }\href
  {\doibase 10.1093/mnras/stab2358} {\bibfield  {journal} {\bibinfo  {journal}
  {Mon. Not. Roy. Astron. Soc.}\ }\textbf {\bibinfo {volume} {507}},\ \bibinfo
  {pages} {5069} (\bibinfo {year} {2021})},\ \Eprint
  {http://arxiv.org/abs/2104.01897} {arXiv:2104.01897 [gr-qc]} \BibitemShut
  {NoStop}%
\bibitem [{\citenamefont {Janquart}\ \emph {et~al.}(2023)\citenamefont
  {Janquart} \emph {et~al.}}]{Janquart:2022fzz}%
  \BibitemOpen
  \bibfield  {author} {\bibinfo {author} {\bibfnamefont {J.}~\bibnamefont
  {Janquart}} \emph {et~al.},\ }\href {\doibase 10.1093/mnras/stad1542}
  {\bibfield  {journal} {\bibinfo  {journal} {Mon. Not. Roy. Astron. Soc.}\
  }\textbf {\bibinfo {volume} {523}},\ \bibinfo {pages} {1699} (\bibinfo {year}
  {2023})},\ \Eprint {http://arxiv.org/abs/2211.01304} {arXiv:2211.01304
  [gr-qc]} \BibitemShut {NoStop}%
\bibitem [{\citenamefont {Relton}\ and\ \citenamefont
  {Raymond}(2021)}]{Relton:2021cax}%
  \BibitemOpen
  \bibfield  {author} {\bibinfo {author} {\bibfnamefont {P.}~\bibnamefont
  {Relton}}\ and\ \bibinfo {author} {\bibfnamefont {V.}~\bibnamefont
  {Raymond}},\ }\href {\doibase 10.1103/PhysRevD.104.084039} {\bibfield
  {journal} {\bibinfo  {journal} {Phys. Rev. D}\ }\textbf {\bibinfo {volume}
  {104}},\ \bibinfo {pages} {084039} (\bibinfo {year} {2021})},\ \Eprint
  {http://arxiv.org/abs/2103.16225} {arXiv:2103.16225 [gr-qc]} \BibitemShut
  {NoStop}%
\bibitem [{\citenamefont {Abbott}\ \emph
  {et~al.}(2020{\natexlab{b}})\citenamefont {Abbott} \emph
  {et~al.}}]{LIGOScientific:2019ryq}%
  \BibitemOpen
  \bibfield  {author} {\bibinfo {author} {\bibfnamefont {B.~P.}\ \bibnamefont
  {Abbott}} \emph {et~al.} (\bibinfo {collaboration} {LIGO Scientific,
  Virgo}),\ }\href {\doibase 10.1103/PhysRevD.101.084002} {\bibfield  {journal}
  {\bibinfo  {journal} {Phys. Rev. D}\ }\textbf {\bibinfo {volume} {101}},\
  \bibinfo {pages} {084002} (\bibinfo {year} {2020}{\natexlab{b}})},\ \Eprint
  {http://arxiv.org/abs/1908.03584} {arXiv:1908.03584 [astro-ph.HE]}
  \BibitemShut {NoStop}%
\bibitem [{\citenamefont {L\'opez~Portilla}\ \emph {et~al.}(2021)\citenamefont
  {L\'opez~Portilla} \emph {et~al.}}]{LopezPortilla:2020odz}%
  \BibitemOpen
  \bibfield  {author} {\bibinfo {author} {\bibfnamefont {M.}~\bibnamefont
  {L\'opez~Portilla}} \emph {et~al.},\ }\href {\doibase
  10.1103/PhysRevD.103.063011} {\bibfield  {journal} {\bibinfo  {journal}
  {Phys. Rev. D}\ }\textbf {\bibinfo {volume} {103}},\ \bibinfo {pages}
  {063011} (\bibinfo {year} {2021})},\ \Eprint
  {http://arxiv.org/abs/2011.13733} {arXiv:2011.13733 [astro-ph.IM]}
  \BibitemShut {NoStop}%
\bibitem [{\citenamefont {Szczepa\'nczyk}\ \emph {et~al.}(2024)\citenamefont
  {Szczepa\'nczyk} \emph {et~al.}}]{Szczepanczyk:2023ihe}%
  \BibitemOpen
  \bibfield  {author} {\bibinfo {author} {\bibfnamefont {M.~J.}\ \bibnamefont
  {Szczepa\'nczyk}} \emph {et~al.},\ }\href {\doibase
  10.1103/PhysRevD.110.042007} {\bibfield  {journal} {\bibinfo  {journal}
  {Phys. Rev. D}\ }\textbf {\bibinfo {volume} {110}},\ \bibinfo {pages}
  {042007} (\bibinfo {year} {2024})},\ \Eprint
  {http://arxiv.org/abs/2305.16146} {arXiv:2305.16146 [astro-ph.HE]}
  \BibitemShut {NoStop}%
\bibitem [{\citenamefont {Dietrich}\ \emph {et~al.}(2021)\citenamefont
  {Dietrich}, \citenamefont {Hinderer},\ and\ \citenamefont
  {Samajdar}}]{Dietrich:2020eud}%
  \BibitemOpen
  \bibfield  {author} {\bibinfo {author} {\bibfnamefont {T.}~\bibnamefont
  {Dietrich}}, \bibinfo {author} {\bibfnamefont {T.}~\bibnamefont {Hinderer}},
  \ and\ \bibinfo {author} {\bibfnamefont {A.}~\bibnamefont {Samajdar}},\
  }\href {\doibase 10.1007/s10714-020-02751-6} {\bibfield  {journal} {\bibinfo
  {journal} {Gen. Rel. Grav.}\ }\textbf {\bibinfo {volume} {53}},\ \bibinfo
  {pages} {27} (\bibinfo {year} {2021})},\ \Eprint
  {http://arxiv.org/abs/2004.02527} {arXiv:2004.02527 [gr-qc]} \BibitemShut
  {NoStop}%
\bibitem [{\citenamefont {Roy}\ and\ \citenamefont
  {Vicente}(2024)}]{Roy:2024rhe}%
  \BibitemOpen
  \bibfield  {author} {\bibinfo {author} {\bibfnamefont {S.}~\bibnamefont
  {Roy}}\ and\ \bibinfo {author} {\bibfnamefont {R.}~\bibnamefont {Vicente}},\
  }\href@noop {} {\  (\bibinfo {year} {2024})},\ \Eprint
  {http://arxiv.org/abs/2410.16388} {arXiv:2410.16388 [gr-qc]} \BibitemShut
  {NoStop}%
\bibitem [{\citenamefont {Leong}\ \emph {et~al.}(2023)\citenamefont {Leong}
  \emph {et~al.}}]{Leong:2023nuk}%
  \BibitemOpen
  \bibfield  {author} {\bibinfo {author} {\bibfnamefont {S.~H.~W.}\
  \bibnamefont {Leong}} \emph {et~al.},\ }\href {\doibase
  10.1103/PhysRevD.108.124079} {\bibfield  {journal} {\bibinfo  {journal}
  {Phys. Rev. D}\ }\textbf {\bibinfo {volume} {108}},\ \bibinfo {pages}
  {124079} (\bibinfo {year} {2023})},\ \Eprint
  {http://arxiv.org/abs/2308.03250} {arXiv:2308.03250 [gr-qc]} \BibitemShut
  {NoStop}%
\bibitem [{\citenamefont {Wright}\ and\ \citenamefont
  {Hendry}(2021)}]{Wright:2021cbn}%
  \BibitemOpen
  \bibfield  {author} {\bibinfo {author} {\bibfnamefont {M.}~\bibnamefont
  {Wright}}\ and\ \bibinfo {author} {\bibfnamefont {M.}~\bibnamefont
  {Hendry}},\ }\href {\doibase 10.3847/1538-4357/ac7ec2} {\  (\bibinfo {year}
  {2021}),\ 10.3847/1538-4357/ac7ec2},\ \Eprint
  {http://arxiv.org/abs/2112.07012} {arXiv:2112.07012 [astro-ph.HE]}
  \BibitemShut {NoStop}%
\bibitem [{\citenamefont {Gupte}\ \emph {et~al.}(2024)\citenamefont {Gupte}
  \emph {et~al.}}]{Gupte:2024jfe}%
  \BibitemOpen
  \bibfield  {author} {\bibinfo {author} {\bibfnamefont {N.}~\bibnamefont
  {Gupte}} \emph {et~al.},\ }\href@noop {} {\  (\bibinfo {year} {2024})},\
  \Eprint {http://arxiv.org/abs/2404.14286} {arXiv:2404.14286 [gr-qc]}
  \BibitemShut {NoStop}%
\bibitem [{\citenamefont {{Einstein Telescope Steering Committee}}()}]{ESFRI}%
  \BibitemOpen
  \bibfield  {author} {\bibinfo {author} {\bibnamefont {{Einstein Telescope
  Steering Committee}}},\ }\href@noop {} {\enquote {\bibinfo {title} {Einstein
  telescope: Science case, design study and feasibility report},}\ }\bibinfo
  {howpublished} {\url{https://apps.et-gw.eu/tds/ql/?c=15662}}\BibitemShut
  {NoStop}%
\bibitem [{\citenamefont {Steltner}\ \emph {et~al.}(2023)\citenamefont
  {Steltner} \emph {et~al.}}]{Steltner:2023cfk}%
  \BibitemOpen
  \bibfield  {author} {\bibinfo {author} {\bibfnamefont {B.}~\bibnamefont
  {Steltner}} \emph {et~al.},\ }\href {\doibase 10.3847/1538-4357/acdad4}
  {\bibfield  {journal} {\bibinfo  {journal} {Astrophys. J.}\ }\textbf
  {\bibinfo {volume} {952}},\ \bibinfo {pages} {55} (\bibinfo {year} {2023})},\
  \Eprint {http://arxiv.org/abs/2303.04109} {arXiv:2303.04109 [gr-qc]}
  \BibitemShut {NoStop}%
\bibitem [{\citenamefont {Steltner}\ \emph {et~al.}(2022)\citenamefont
  {Steltner}, \citenamefont {Papa},\ and\ \citenamefont
  {Eggenstein}}]{Steltner:2021qjy}%
  \BibitemOpen
  \bibfield  {author} {\bibinfo {author} {\bibfnamefont {B.}~\bibnamefont
  {Steltner}}, \bibinfo {author} {\bibfnamefont {M.~A.}\ \bibnamefont {Papa}},
  \ and\ \bibinfo {author} {\bibfnamefont {H.~B.}\ \bibnamefont {Eggenstein}},\
  }\href {\doibase 10.1103/PhysRevD.105.022005} {\bibfield  {journal} {\bibinfo
   {journal} {Phys. Rev. D}\ }\textbf {\bibinfo {volume} {105}},\ \bibinfo
  {pages} {022005} (\bibinfo {year} {2022})},\ \Eprint
  {http://arxiv.org/abs/2105.09933} {arXiv:2105.09933 [gr-qc]} \BibitemShut
  {NoStop}%
\bibitem [{\citenamefont {Davis}\ \emph {et~al.}(2019)\citenamefont {Davis}
  \emph {et~al.}}]{Davis:2018yrz}%
  \BibitemOpen
  \bibfield  {author} {\bibinfo {author} {\bibfnamefont {D.}~\bibnamefont
  {Davis}} \emph {et~al.},\ }\href {\doibase 10.1088/1361-6382/ab01c5}
  {\bibfield  {journal} {\bibinfo  {journal} {Class. Quant. Grav.}\ }\textbf
  {\bibinfo {volume} {36}},\ \bibinfo {pages} {055011} (\bibinfo {year}
  {2019})},\ \Eprint {http://arxiv.org/abs/1809.05348} {arXiv:1809.05348
  [astro-ph.IM]} \BibitemShut {NoStop}%
\bibitem [{\citenamefont {Driggers}\ \emph {et~al.}(2019)\citenamefont
  {Driggers} \emph {et~al.}}]{LIGOScientific:2018kdd}%
  \BibitemOpen
  \bibfield  {author} {\bibinfo {author} {\bibfnamefont {J.~C.}\ \bibnamefont
  {Driggers}} \emph {et~al.} (\bibinfo {collaboration} {LIGO Scientific}),\
  }\href {\doibase 10.1103/PhysRevD.99.042001} {\bibfield  {journal} {\bibinfo
  {journal} {Phys. Rev. D}\ }\textbf {\bibinfo {volume} {99}},\ \bibinfo
  {pages} {042001} (\bibinfo {year} {2019})},\ \Eprint
  {http://arxiv.org/abs/1806.00532} {arXiv:1806.00532 [astro-ph.IM]}
  \BibitemShut {NoStop}%
\bibitem [{\citenamefont {Lopez}\ \emph
  {et~al.}(2022{\natexlab{a}})\citenamefont {Lopez} \emph
  {et~al.}}]{Lopez:2022lkd}%
  \BibitemOpen
  \bibfield  {author} {\bibinfo {author} {\bibfnamefont {M.}~\bibnamefont
  {Lopez}} \emph {et~al.},\ }\href {\doibase 10.1103/PhysRevD.106.023027}
  {\bibfield  {journal} {\bibinfo  {journal} {Phys. Rev. D}\ }\textbf {\bibinfo
  {volume} {106}},\ \bibinfo {pages} {023027} (\bibinfo {year}
  {2022}{\natexlab{a}})},\ \Eprint {http://arxiv.org/abs/2203.06494}
  {arXiv:2203.06494 [astro-ph.IM]} \BibitemShut {NoStop}%
\bibitem [{\citenamefont {Lopez}\ \emph
  {et~al.}(2022{\natexlab{b}})\citenamefont {Lopez} \emph
  {et~al.}}]{Lopez:2022dho}%
  \BibitemOpen
  \bibfield  {author} {\bibinfo {author} {\bibfnamefont {M.}~\bibnamefont
  {Lopez}} \emph {et~al.},\ }\href@noop {} {\  (\bibinfo {year}
  {2022}{\natexlab{b}})},\ \Eprint {http://arxiv.org/abs/2205.09204}
  {arXiv:2205.09204 [astro-ph.IM]} \BibitemShut {NoStop}%
\bibitem [{\citenamefont {Guersel}\ and\ \citenamefont
  {Tinto}(1989)}]{Guersel:1989th}%
  \BibitemOpen
  \bibfield  {author} {\bibinfo {author} {\bibfnamefont {Y.}~\bibnamefont
  {Guersel}}\ and\ \bibinfo {author} {\bibfnamefont {M.}~\bibnamefont
  {Tinto}},\ }\href {\doibase 10.1103/PhysRevD.40.3884} {\bibfield  {journal}
  {\bibinfo  {journal} {Phys. Rev. D}\ }\textbf {\bibinfo {volume} {40}},\
  \bibinfo {pages} {3884} (\bibinfo {year} {1989})}\BibitemShut {NoStop}%
\bibitem [{\citenamefont {Freise}\ \emph {et~al.}(2009)\citenamefont {Freise}
  \emph {et~al.}}]{Freise:2008dk}%
  \BibitemOpen
  \bibfield  {author} {\bibinfo {author} {\bibfnamefont {A.}~\bibnamefont
  {Freise}} \emph {et~al.},\ }\href {\doibase 10.1088/0264-9381/26/8/085012}
  {\bibfield  {journal} {\bibinfo  {journal} {Class. Quant. Grav.}\ }\textbf
  {\bibinfo {volume} {26}},\ \bibinfo {pages} {085012} (\bibinfo {year}
  {2009})},\ \Eprint {http://arxiv.org/abs/0804.1036} {arXiv:0804.1036 [gr-qc]}
  \BibitemShut {NoStop}%
\bibitem [{\citenamefont {Sutton}\ \emph {et~al.}(2010)\citenamefont {Sutton}
  \emph {et~al.}}]{Sutton:2009gi}%
  \BibitemOpen
  \bibfield  {author} {\bibinfo {author} {\bibfnamefont {P.~J.}\ \bibnamefont
  {Sutton}} \emph {et~al.},\ }\href {\doibase 10.1088/1367-2630/12/5/053034}
  {\bibfield  {journal} {\bibinfo  {journal} {New J. Phys.}\ }\textbf {\bibinfo
  {volume} {12}},\ \bibinfo {pages} {053034} (\bibinfo {year} {2010})},\
  \Eprint {http://arxiv.org/abs/0908.3665} {arXiv:0908.3665 [gr-qc]}
  \BibitemShut {NoStop}%
\bibitem [{\citenamefont {Pang}\ \emph {et~al.}(2020)\citenamefont {Pang} \emph
  {et~al.}}]{Pang:2020pfz}%
  \BibitemOpen
  \bibfield  {author} {\bibinfo {author} {\bibfnamefont {P.~T.~H.}\
  \bibnamefont {Pang}} \emph {et~al.},\ }\href {\doibase
  10.1103/PhysRevD.101.104055} {\bibfield  {journal} {\bibinfo  {journal}
  {Phys. Rev. D}\ }\textbf {\bibinfo {volume} {101}},\ \bibinfo {pages}
  {104055} (\bibinfo {year} {2020})},\ \Eprint
  {http://arxiv.org/abs/2003.07375} {arXiv:2003.07375 [gr-qc]} \BibitemShut
  {NoStop}%
\bibitem [{\citenamefont {Wong}\ \emph {et~al.}(2021)\citenamefont {Wong} \emph
  {et~al.}}]{Wong:2021cmp}%
  \BibitemOpen
  \bibfield  {author} {\bibinfo {author} {\bibfnamefont {I.~C.~F.}\
  \bibnamefont {Wong}} \emph {et~al.},\ }\href@noop {} {\  (\bibinfo {year}
  {2021})},\ \Eprint {http://arxiv.org/abs/2105.09485} {arXiv:2105.09485
  [gr-qc]} \BibitemShut {NoStop}%
\bibitem [{\citenamefont {Goncharov}\ \emph {et~al.}(2022)\citenamefont
  {Goncharov}, \citenamefont {Nitz},\ and\ \citenamefont
  {Harms}}]{Goncharov:2022dgl}%
  \BibitemOpen
  \bibfield  {author} {\bibinfo {author} {\bibfnamefont {B.}~\bibnamefont
  {Goncharov}}, \bibinfo {author} {\bibfnamefont {A.~H.}\ \bibnamefont {Nitz}},
  \ and\ \bibinfo {author} {\bibfnamefont {J.}~\bibnamefont {Harms}},\ }\href
  {\doibase 10.1103/PhysRevD.105.122007} {\bibfield  {journal} {\bibinfo
  {journal} {Phys. Rev. D}\ }\textbf {\bibinfo {volume} {105}},\ \bibinfo
  {pages} {122007} (\bibinfo {year} {2022})},\ \Eprint
  {http://arxiv.org/abs/2204.08533} {arXiv:2204.08533 [gr-qc]} \BibitemShut
  {NoStop}%
\bibitem [{\citenamefont {Cornish}\ and\ \citenamefont
  {Littenberg}(2015)}]{Cornish:2014kda}%
  \BibitemOpen
  \bibfield  {author} {\bibinfo {author} {\bibfnamefont {N.~J.}\ \bibnamefont
  {Cornish}}\ and\ \bibinfo {author} {\bibfnamefont {T.~B.}\ \bibnamefont
  {Littenberg}},\ }\href {\doibase 10.1088/0264-9381/32/13/135012} {\bibfield
  {journal} {\bibinfo  {journal} {Class. Quant. Grav.}\ }\textbf {\bibinfo
  {volume} {32}},\ \bibinfo {pages} {135012} (\bibinfo {year} {2015})},\
  \Eprint {http://arxiv.org/abs/1410.3835} {arXiv:1410.3835 [gr-qc]}
  \BibitemShut {NoStop}%
\bibitem [{\citenamefont {Cornish}\ \emph {et~al.}(2021)\citenamefont {Cornish}
  \emph {et~al.}}]{Cornish:2020dwh}%
  \BibitemOpen
  \bibfield  {author} {\bibinfo {author} {\bibfnamefont {N.~J.}\ \bibnamefont
  {Cornish}} \emph {et~al.},\ }\href {\doibase 10.1103/PhysRevD.103.044006}
  {\bibfield  {journal} {\bibinfo  {journal} {Phys. Rev. D}\ }\textbf {\bibinfo
  {volume} {103}},\ \bibinfo {pages} {044006} (\bibinfo {year} {2021})},\
  \Eprint {http://arxiv.org/abs/2011.09494} {arXiv:2011.09494 [gr-qc]}
  \BibitemShut {NoStop}%
\bibitem [{\citenamefont {Veitch}\ and\ \citenamefont
  {Vecchio}(2010)}]{Veitch:2009hd}%
  \BibitemOpen
  \bibfield  {author} {\bibinfo {author} {\bibfnamefont {J.}~\bibnamefont
  {Veitch}}\ and\ \bibinfo {author} {\bibfnamefont {A.}~\bibnamefont
  {Vecchio}},\ }\href {\doibase 10.1103/PhysRevD.81.062003} {\bibfield
  {journal} {\bibinfo  {journal} {Phys. Rev. D}\ }\textbf {\bibinfo {volume}
  {81}},\ \bibinfo {pages} {062003} (\bibinfo {year} {2010})},\ \Eprint
  {http://arxiv.org/abs/0911.3820} {arXiv:0911.3820 [astro-ph.CO]} \BibitemShut
  {NoStop}%
\bibitem [{\citenamefont {Green}(1995)}]{Green:1995mxx}%
  \BibitemOpen
  \bibfield  {author} {\bibinfo {author} {\bibfnamefont {P.~J.}\ \bibnamefont
  {Green}},\ }\href {\doibase 10.1093/biomet/82.4.711} {\bibfield  {journal}
  {\bibinfo  {journal} {Biometrika}\ }\textbf {\bibinfo {volume} {82}},\
  \bibinfo {pages} {711} (\bibinfo {year} {1995})}\BibitemShut {NoStop}%
\bibitem [{\citenamefont {Khan}\ \emph {et~al.}(2016)\citenamefont {Khan} \emph
  {et~al.}}]{Khan:2015jqa}%
  \BibitemOpen
  \bibfield  {author} {\bibinfo {author} {\bibfnamefont {S.}~\bibnamefont
  {Khan}} \emph {et~al.},\ }\href {\doibase 10.1103/PhysRevD.93.044007}
  {\bibfield  {journal} {\bibinfo  {journal} {Phys. Rev. D}\ }\textbf {\bibinfo
  {volume} {93}},\ \bibinfo {pages} {044007} (\bibinfo {year} {2016})},\
  \Eprint {http://arxiv.org/abs/1508.07253} {arXiv:1508.07253 [gr-qc]}
  \BibitemShut {NoStop}%
\bibitem [{\citenamefont {Belczynski}\ \emph {et~al.}(2017)\citenamefont
  {Belczynski} \emph {et~al.}}]{Belczynski:2016ieo}%
  \BibitemOpen
  \bibfield  {author} {\bibinfo {author} {\bibfnamefont {K.}~\bibnamefont
  {Belczynski}} \emph {et~al.},\ }\href {\doibase 10.1093/mnras/stx1759}
  {\bibfield  {journal} {\bibinfo  {journal} {Mon. Not. Roy. Astron. Soc.}\
  }\textbf {\bibinfo {volume} {471}},\ \bibinfo {pages} {4702} (\bibinfo {year}
  {2017})},\ \Eprint {http://arxiv.org/abs/1612.01524} {arXiv:1612.01524
  [astro-ph.HE]} \BibitemShut {NoStop}%
\bibitem [{\citenamefont {Oguri}(2018)}]{Oguri:2018muv}%
  \BibitemOpen
  \bibfield  {author} {\bibinfo {author} {\bibfnamefont {M.}~\bibnamefont
  {Oguri}},\ }\href {\doibase 10.1093/mnras/sty2145} {\bibfield  {journal}
  {\bibinfo  {journal} {Mon. Not. Roy. Astron. Soc.}\ }\textbf {\bibinfo
  {volume} {480}},\ \bibinfo {pages} {3842} (\bibinfo {year} {2018})},\ \Eprint
  {http://arxiv.org/abs/1807.02584} {arXiv:1807.02584 [astro-ph.CO]}
  \BibitemShut {NoStop}%
\bibitem [{\citenamefont {Lopez}()}]{Lopez:gengli_url}%
  \BibitemOpen
  \bibfield  {author} {\bibinfo {author} {\bibfnamefont {M.}~\bibnamefont
  {Lopez}},\ }\href@noop {} {\enquote {\bibinfo {title} {gengli’s
  documentation},}\ }\bibinfo {howpublished}
  {\url{https://melissa.lopez.docs.ligo.org/gengli/index.htmll}},\ \bibinfo
  {note} {[Accessed 06-11-2024]}\BibitemShut {NoStop}%
\bibitem [{\citenamefont {Schutz}(1986)}]{Schutz:1986gp}%
  \BibitemOpen
  \bibfield  {author} {\bibinfo {author} {\bibfnamefont {B.~F.}\ \bibnamefont
  {Schutz}},\ }\href {\doibase 10.1038/323310a0} {\bibfield  {journal}
  {\bibinfo  {journal} {Nature}\ }\textbf {\bibinfo {volume} {323}},\ \bibinfo
  {pages} {310} (\bibinfo {year} {1986})}\BibitemShut {NoStop}%
\bibitem [{\citenamefont {MacLeod}\ and\ \citenamefont
  {Hogan}(2008)}]{MacLeod:2007jd}%
  \BibitemOpen
  \bibfield  {author} {\bibinfo {author} {\bibfnamefont {C.~L.}\ \bibnamefont
  {MacLeod}}\ and\ \bibinfo {author} {\bibfnamefont {C.~J.}\ \bibnamefont
  {Hogan}},\ }\href {\doibase 10.1103/PhysRevD.77.043512} {\bibfield  {journal}
  {\bibinfo  {journal} {Phys. Rev. D}\ }\textbf {\bibinfo {volume} {77}},\
  \bibinfo {pages} {043512} (\bibinfo {year} {2008})},\ \Eprint
  {http://arxiv.org/abs/0712.0618} {arXiv:0712.0618 [astro-ph]} \BibitemShut
  {NoStop}%
\bibitem [{\citenamefont {Del~Pozzo}(2012)}]{DelPozzo:2011vcw}%
  \BibitemOpen
  \bibfield  {author} {\bibinfo {author} {\bibfnamefont {W.}~\bibnamefont
  {Del~Pozzo}},\ }\href {\doibase 10.1103/PhysRevD.86.043011} {\bibfield
  {journal} {\bibinfo  {journal} {Phys. Rev. D}\ }\textbf {\bibinfo {volume}
  {86}},\ \bibinfo {pages} {043011} (\bibinfo {year} {2012})},\ \Eprint
  {http://arxiv.org/abs/1108.1317} {arXiv:1108.1317 [astro-ph.CO]} \BibitemShut
  {NoStop}%
\bibitem [{\citenamefont {Chen}\ \emph {et~al.}(2018)\citenamefont {Chen},
  \citenamefont {Fishbach},\ and\ \citenamefont {Holz}}]{Chen:2017rfc}%
  \BibitemOpen
  \bibfield  {author} {\bibinfo {author} {\bibfnamefont {H.-Y.}\ \bibnamefont
  {Chen}}, \bibinfo {author} {\bibfnamefont {M.}~\bibnamefont {Fishbach}}, \
  and\ \bibinfo {author} {\bibfnamefont {D.~E.}\ \bibnamefont {Holz}},\ }\href
  {\doibase 10.1038/s41586-018-0606-0} {\bibfield  {journal} {\bibinfo
  {journal} {Nature}\ }\textbf {\bibinfo {volume} {562}},\ \bibinfo {pages}
  {545} (\bibinfo {year} {2018})},\ \Eprint {http://arxiv.org/abs/1712.06531}
  {arXiv:1712.06531 [astro-ph.CO]} \BibitemShut {NoStop}%
\bibitem [{\citenamefont {Fishbach}\ \emph {et~al.}(2019)\citenamefont
  {Fishbach} \emph {et~al.}}]{LIGOScientific:2018gmd}%
  \BibitemOpen
  \bibfield  {author} {\bibinfo {author} {\bibfnamefont {M.}~\bibnamefont
  {Fishbach}} \emph {et~al.} (\bibinfo {collaboration} {LIGO Scientific,
  Virgo}),\ }\href {\doibase 10.3847/2041-8213/aaf96e} {\bibfield  {journal}
  {\bibinfo  {journal} {Astrophys. J. Lett.}\ }\textbf {\bibinfo {volume}
  {871}},\ \bibinfo {pages} {L13} (\bibinfo {year} {2019})},\ \Eprint
  {http://arxiv.org/abs/1807.05667} {arXiv:1807.05667 [astro-ph.CO]}
  \BibitemShut {NoStop}%
\bibitem [{\citenamefont {Gray}\ \emph {et~al.}(2020)\citenamefont {Gray} \emph
  {et~al.}}]{Gray:2019ksv}%
  \BibitemOpen
  \bibfield  {author} {\bibinfo {author} {\bibfnamefont {R.}~\bibnamefont
  {Gray}} \emph {et~al.},\ }\href {\doibase 10.1103/PhysRevD.101.122001}
  {\bibfield  {journal} {\bibinfo  {journal} {Phys. Rev. D}\ }\textbf {\bibinfo
  {volume} {101}},\ \bibinfo {pages} {122001} (\bibinfo {year} {2020})},\
  \Eprint {http://arxiv.org/abs/1908.06050} {arXiv:1908.06050 [gr-qc]}
  \BibitemShut {NoStop}%
\bibitem [{\citenamefont {Soares-Santos}\ \emph {et~al.}(2019)\citenamefont
  {Soares-Santos} \emph {et~al.}}]{DES:2019ccw}%
  \BibitemOpen
  \bibfield  {author} {\bibinfo {author} {\bibfnamefont {M.}~\bibnamefont
  {Soares-Santos}} \emph {et~al.} (\bibinfo {collaboration} {DES, LIGO
  Scientific, Virgo}),\ }\href {\doibase 10.3847/2041-8213/ab14f1} {\bibfield
  {journal} {\bibinfo  {journal} {Astrophys. J. Lett.}\ }\textbf {\bibinfo
  {volume} {876}},\ \bibinfo {pages} {L7} (\bibinfo {year} {2019})},\ \Eprint
  {http://arxiv.org/abs/1901.01540} {arXiv:1901.01540 [astro-ph.CO]}
  \BibitemShut {NoStop}%
\bibitem [{\citenamefont {Palmese}\ \emph {et~al.}(2020)\citenamefont {Palmese}
  \emph {et~al.}}]{DES:2020nay}%
  \BibitemOpen
  \bibfield  {author} {\bibinfo {author} {\bibfnamefont {A.}~\bibnamefont
  {Palmese}} \emph {et~al.} (\bibinfo {collaboration} {DES}),\ }\href {\doibase
  10.3847/2041-8213/abaeff} {\bibfield  {journal} {\bibinfo  {journal}
  {Astrophys. J. Lett.}\ }\textbf {\bibinfo {volume} {900}},\ \bibinfo {pages}
  {L33} (\bibinfo {year} {2020})},\ \Eprint {http://arxiv.org/abs/2006.14961}
  {arXiv:2006.14961 [astro-ph.CO]} \BibitemShut {NoStop}%
\bibitem [{\citenamefont {Abbott}\ \emph
  {et~al.}(2021{\natexlab{d}})\citenamefont {Abbott} \emph
  {et~al.}}]{LIGOScientific:2019zcs}%
  \BibitemOpen
  \bibfield  {author} {\bibinfo {author} {\bibfnamefont {B.~P.}\ \bibnamefont
  {Abbott}} \emph {et~al.} (\bibinfo {collaboration} {LIGO Scientific, Virgo,
  VIRGO}),\ }\href {\doibase 10.3847/1538-4357/abdcb7} {\bibfield  {journal}
  {\bibinfo  {journal} {Astrophys. J.}\ }\textbf {\bibinfo {volume} {909}},\
  \bibinfo {pages} {218} (\bibinfo {year} {2021}{\natexlab{d}})},\ \Eprint
  {http://arxiv.org/abs/1908.06060} {arXiv:1908.06060 [astro-ph.CO]}
  \BibitemShut {NoStop}%
\bibitem [{\citenamefont {Finke}\ \emph {et~al.}(2021)\citenamefont {Finke},
  \citenamefont {Foffa}, \citenamefont {Iacovelli}, \citenamefont {Maggiore},\
  and\ \citenamefont {Mancarella}}]{Finke:2021aom}%
  \BibitemOpen
  \bibfield  {author} {\bibinfo {author} {\bibfnamefont {A.}~\bibnamefont
  {Finke}}, \bibinfo {author} {\bibfnamefont {S.}~\bibnamefont {Foffa}},
  \bibinfo {author} {\bibfnamefont {F.}~\bibnamefont {Iacovelli}}, \bibinfo
  {author} {\bibfnamefont {M.}~\bibnamefont {Maggiore}}, \ and\ \bibinfo
  {author} {\bibfnamefont {M.}~\bibnamefont {Mancarella}},\ }\href {\doibase
  10.1088/1475-7516/2021/08/026} {\bibfield  {journal} {\bibinfo  {journal}
  {JCAP}\ }\textbf {\bibinfo {volume} {08}},\ \bibinfo {pages} {026} (\bibinfo
  {year} {2021})},\ \Eprint {http://arxiv.org/abs/2101.12660} {arXiv:2101.12660
  [astro-ph.CO]} \BibitemShut {NoStop}%
\bibitem [{\citenamefont {Palmese}\ \emph {et~al.}(2023)\citenamefont
  {Palmese}, \citenamefont {Bom}, \citenamefont {Mucesh},\ and\ \citenamefont
  {Hartley}}]{Palmese:2021mjm}%
  \BibitemOpen
  \bibfield  {author} {\bibinfo {author} {\bibfnamefont {A.}~\bibnamefont
  {Palmese}}, \bibinfo {author} {\bibfnamefont {C.~R.}\ \bibnamefont {Bom}},
  \bibinfo {author} {\bibfnamefont {S.}~\bibnamefont {Mucesh}}, \ and\ \bibinfo
  {author} {\bibfnamefont {W.~G.}\ \bibnamefont {Hartley}},\ }\href {\doibase
  10.3847/1538-4357/aca6e3} {\bibfield  {journal} {\bibinfo  {journal}
  {Astrophys. J.}\ }\textbf {\bibinfo {volume} {943}},\ \bibinfo {pages} {56}
  (\bibinfo {year} {2023})},\ \Eprint {http://arxiv.org/abs/2111.06445}
  {arXiv:2111.06445 [astro-ph.CO]} \BibitemShut {NoStop}%
\bibitem [{\citenamefont {Abbott}\ \emph
  {et~al.}(2017{\natexlab{b}})\citenamefont {Abbott} \emph
  {et~al.}}]{LIGOScientific:2017bnn}%
  \BibitemOpen
  \bibfield  {author} {\bibinfo {author} {\bibfnamefont {B.~P.}\ \bibnamefont
  {Abbott}} \emph {et~al.} (\bibinfo {collaboration} {LIGO Scientific,
  VIRGO}),\ }\href {\doibase 10.1103/PhysRevLett.118.221101} {\bibfield
  {journal} {\bibinfo  {journal} {Phys. Rev. Lett.}\ }\textbf {\bibinfo
  {volume} {118}},\ \bibinfo {pages} {221101} (\bibinfo {year}
  {2017}{\natexlab{b}})},\ \bibinfo {note} {[Erratum: Phys.Rev.Lett. 121,
  129901 (2018)]},\ \Eprint {http://arxiv.org/abs/1706.01812} {arXiv:1706.01812
  [gr-qc]} \BibitemShut {NoStop}%
\bibitem [{\citenamefont {Abbott}\ \emph
  {et~al.}(2019{\natexlab{b}})\citenamefont {Abbott} \emph
  {et~al.}}]{LIGOScientific:2018dkp}%
  \BibitemOpen
  \bibfield  {author} {\bibinfo {author} {\bibfnamefont {B.~P.}\ \bibnamefont
  {Abbott}} \emph {et~al.} (\bibinfo {collaboration} {LIGO Scientific,
  Virgo}),\ }\href {\doibase 10.1103/PhysRevLett.123.011102} {\bibfield
  {journal} {\bibinfo  {journal} {Phys. Rev. Lett.}\ }\textbf {\bibinfo
  {volume} {123}},\ \bibinfo {pages} {011102} (\bibinfo {year}
  {2019}{\natexlab{b}})},\ \Eprint {http://arxiv.org/abs/1811.00364}
  {arXiv:1811.00364 [gr-qc]} \BibitemShut {NoStop}%
\bibitem [{\citenamefont {Narola}\ \emph {et~al.}(2024)\citenamefont {Narola},
  \citenamefont {Janquart}, \citenamefont {Haegel}, \citenamefont {Haris},
  \citenamefont {Hannuksela},\ and\ \citenamefont {Van
  Den~Broeck}}]{Narola:2023viz}%
  \BibitemOpen
  \bibfield  {author} {\bibinfo {author} {\bibfnamefont {H.}~\bibnamefont
  {Narola}}, \bibinfo {author} {\bibfnamefont {J.}~\bibnamefont {Janquart}},
  \bibinfo {author} {\bibfnamefont {L.}~\bibnamefont {Haegel}}, \bibinfo
  {author} {\bibfnamefont {K.}~\bibnamefont {Haris}}, \bibinfo {author}
  {\bibfnamefont {O.~A.}\ \bibnamefont {Hannuksela}}, \ and\ \bibinfo {author}
  {\bibfnamefont {C.}~\bibnamefont {Van Den~Broeck}},\ }\href {\doibase
  10.1103/PhysRevD.109.084064} {\bibfield  {journal} {\bibinfo  {journal}
  {Phys. Rev. D}\ }\textbf {\bibinfo {volume} {109}},\ \bibinfo {pages}
  {084064} (\bibinfo {year} {2024})},\ \Eprint
  {http://arxiv.org/abs/2308.01709} {arXiv:2308.01709 [gr-qc]} \BibitemShut
  {NoStop}%
\bibitem [{\citenamefont {Gupta}\ \emph
  {et~al.}(2024{\natexlab{b}})\citenamefont {Gupta} \emph
  {et~al.}}]{Gupta:2023lga}%
  \BibitemOpen
  \bibfield  {author} {\bibinfo {author} {\bibfnamefont {I.}~\bibnamefont
  {Gupta}} \emph {et~al.},\ }\href {\doibase 10.1088/1361-6382/ad7b99}
  {\bibfield  {journal} {\bibinfo  {journal} {Class. Quant. Grav.}\ }\textbf
  {\bibinfo {volume} {41}},\ \bibinfo {pages} {245001} (\bibinfo {year}
  {2024}{\natexlab{b}})},\ \Eprint {http://arxiv.org/abs/2307.10421}
  {arXiv:2307.10421 [gr-qc]} \BibitemShut {NoStop}%
\bibitem [{\citenamefont {Nitz}\ and\ \citenamefont
  {Dal~Canton}(2021)}]{Nitz:2021pbr}%
  \BibitemOpen
  \bibfield  {author} {\bibinfo {author} {\bibfnamefont {A.~H.}\ \bibnamefont
  {Nitz}}\ and\ \bibinfo {author} {\bibfnamefont {T.}~\bibnamefont
  {Dal~Canton}},\ }\href {\doibase 10.3847/2041-8213/ac1a75} {\bibfield
  {journal} {\bibinfo  {journal} {Astrophys. J. Lett.}\ }\textbf {\bibinfo
  {volume} {917}},\ \bibinfo {pages} {L27} (\bibinfo {year} {2021})},\ \Eprint
  {http://arxiv.org/abs/2106.15259} {arXiv:2106.15259 [astro-ph.HE]}
  \BibitemShut {NoStop}%
\bibitem [{\citenamefont {Abdikamalov}\ \emph {et~al.}(2021)\citenamefont
  {Abdikamalov}, \citenamefont {Pagliaroli},\ and\ \citenamefont
  {Radice}}]{Abdikamalov_2021}%
  \BibitemOpen
  \bibfield  {author} {\bibinfo {author} {\bibfnamefont {E.}~\bibnamefont
  {Abdikamalov}}, \bibinfo {author} {\bibfnamefont {G.}~\bibnamefont
  {Pagliaroli}}, \ and\ \bibinfo {author} {\bibfnamefont {D.}~\bibnamefont
  {Radice}},\ }\enquote {\bibinfo {title} {Gravitational waves from
  core-collapse supernovae},}\ in\ \href {\doibase
  10.1007/978-981-15-4702-7_21-1} {\emph {\bibinfo {booktitle} {Handbook of
  Gravitational Wave Astronomy}}}\ (\bibinfo  {publisher} {Springer
  Singapore},\ \bibinfo {year} {2021})\ p.\ \bibinfo {pages}
  {1–37}\BibitemShut {NoStop}%
\bibitem [{\citenamefont {Abac}\ \emph {et~al.}(2025)\citenamefont {Abac} \emph
  {et~al.}}]{Abac:2025saz}%
  \BibitemOpen
  \bibfield  {author} {\bibinfo {author} {\bibfnamefont {A.}~\bibnamefont
  {Abac}} \emph {et~al.},\ }\href@noop {} {\  (\bibinfo {year} {2025})},\
  \Eprint {http://arxiv.org/abs/2503.12263} {arXiv:2503.12263 [gr-qc]}
  \BibitemShut {NoStop}%
\end{thebibliography}%
\end{document}